\documentclass[a4paper]{aa}
\usepackage[varg]{txfonts}
\pdfoutput=1

\usepackage{hyperref}
\hypersetup{colorlinks=true,linkcolor=blue,citecolor=blue,filecolor=blue,urlcolor=blue}

\usepackage{graphicx}
\graphicspath{{images/}}

\usepackage{bm}

\newcommand{\pc}{\mathrm{pc}}
\newcommand{\magn}{\mathrm{mag}}
\newcommand{\myr}{\mathrm{Myr}}
\newcommand{\msun}{M_\odot}

\newcommand{\adeg}{^\circ}
\newcommand{\amin}{^\prime}

\newcommand{\mcl}{M_\mathrm{cl}}

\newcommand{\rh}{r_\mathrm{h}}
\newcommand{\mseg}{\texttt{seg}}
\newcommand{\mnon}{\texttt{non}}
\newcommand{\trel}{t_\mathrm{rh}}
\newcommand{\tcr}{t_\mathrm{cr}}
\newcommand{\tauv}{\tau_\mathrm{v}}
\newcommand{\nbin}{n_\mathrm{bin}}

\begin{document}

\title{Do star clusters form in a completely mass-segregated way?\thanks{Table \ref{tab:data} is only available in electronic form at the CDS via anonymous ftp to cdsarc.u-strasbg.fr (130.79.128.5) or via \url{http://cdsweb.u-strasbg.fr/cgi-bin/qcat?J/A+A/}}}
\subtitle{}

\titlerunning{Do star clusters form in a completely mass-segregated way?}        

\setcounter{footnote}{4}

\author{Václav Pavlík\inst{\ref{auuk},\ref{obs},}\thanks{\email{pavlik@sirrah.troja.mff.cuni.cz}}
\and Pavel Kroupa\inst{\ref{auuk},\ref{bonn}}
\and Ladislav Šubr\inst{\ref{auuk}}}

\institute{Astronomical Institute of Charles University, Prague, Czech Republic\label{auuk}
\and Observatory and Planetarium of Prague, Prague, Czech Republic\label{obs}
\and Helmholtz-Institut für Strahlen- und Kernphysik (HISKP), Universität Bonn, Bonn, Germany\label{bonn}}

\authorrunning{Pavlík et al.}

\date{Received: September 18, 2018 / Accepted: May 13, 2019}

\abstract
{ALMA observations of the Serpens South star-forming region suggest that stellar protoclusters may be completely mass segregated at birth. Independent observations also suggest that embedded clusters form segregated by mass.}
{As the primordial mass segregation seems to be lost over time, we aim to study on which timescale an initially perfectly mass-segregated star cluster becomes indistinguishable from an initially not mass-segregated cluster. As an example, the Orion Nebula Cluster (ONC) is also discussed.}
{We used $N$-body simulations of star clusters with various masses and two different degrees of primordial mass segregation. We analysed their energy redistribution through two-body relaxation to quantify the time when the models agree in terms of mass segregation, which sets in only dynamically in the models that are primordially not mass segregated. A comprehensive cross-matched catalogue combining optical, infrared, and X-ray surveys of ONC members was also compiled and made available.}
{The models evolve to a similar radial distribution of high-mass stars after the core collapse (about half a median two-body relaxation time, $\trel$) and become observationally indistinguishable from the point of view of mass segregation at time $\tauv \approx 3.3\,\trel\,$. In the case of the ONC, using the distribution of high-mass stars, we may not rule out either evolutionary scenario (regardless of whether they are initially mass segregated). When we account for extinction and elongation of the ONC, as reported elsewhere, an initially perfectly mass-segregated state seems to be more consistent with the observed cluster.}
{}


\maketitle

\section{Introduction}
\label{sec:intro}

According to gravothermal turbulence \citep{pad_nor02,hen_chab08,hen_chab11}, star formation is largely a random process taking place in stochastically occurring density peaks in molecular clouds. Observations with the \emph{Herschel} space telescope and with the \emph{ALMA} facility have shown, however, that stars form in thin and long ($\approx0.1\,\pc$ wide and a few to many parsec long) kinematically coherent filaments of molecular material \citep{andre_etal2010,andre_etal,hacar_etal1,hacar_etal2,mattern18}, suggesting instead that the formation of stars occurs in a non-turbulent and more constrained environment. In addition, if protostars throttle their accretion through the accretion-induced luminosity \citep{ada_fat96,mat_mck00}, then a star's mass depends on the accretion rate and thus on the inflow rate from the filament. A low-mass cloud core yielding a low accretion rate will terminate accretion sooner because a lower feedback energy will be required. Denser and more massive cores yielding higher accretion rates will allow the star to grow to a higher mass until the feedback energy can oppose the infall. In this case, star formation would be highly self-regulated. This scenario is naturally consistent with the relation $m_\mathrm{max} - \mcl$ (most massive star, stellar-cluster mass, respectively) observed in very young clusters \citep{weidner10,weidner13,kirk_myers11,stephens17,ramirez16}.

One possible implication of these theoretical ideas is that embedded clusters may be less (if their formation is not self-regulated) or more segregated by mass primordially (if their formation is self-regulated). Studying primordial mass segregation is thus relevant for understanding the process of star formation. It is also important for understanding the evolution of star clusters because initially mass-segregated clusters tend to more quickly loose their low-mass stars across their tidal boundary when residual gas expulsion plays a role \citep{haghi_etal15}. The mass-segregated clusters will eject their massive stars more quickly and efficiently as a result of the energetic encounters between massive stars in the cluster cores \citep{oh_etal15,oh_kroupa16,kroupa_onc,long_etal18}. Primordial mass segregation also affects the dynamical evolution of a cluster through stellar evolutionary mass loss of the massive stars. Initial mass segregation may also lead to a different evolution of the radial dependence of the stellar mass function in the evolving star cluster when compared to a cluster not initially mass segregated \citep{webb_vesp}.

It is therefore of interest to understand whether embedded star clusters are born segregated by mass, and how strong the mass segregation is. The stars in a star cluster in dynamical equilibrium that is completely mass segregated initially\footnote{This does not mean, however, that such a system is also in a state of energy equipartition.} are almost perfectly arranged by mass and energy from the centre of the cluster outward \citep{baumgardt_segr}; another method for creating mass-segregated clusters, also developed in Bonn, uses interparticle potentials and binding energy to sort the stars \hbox{\citep{subr_model}}.
Both methods imply\footnote{The former strictly, the latter statistically.} that the most massive stars have the lowest energy (i.e.\ they are confined to the centre), while the least massive stars have the highest energy (i.e.\ they are on the widest orbits and are initially placed at the largest distance from the centre). This is consistent with the notion that the most massive stars in an embedded cluster need the highest densities to form \citep{vazquez_etal}.
A star cluster that is not primordially mass segregated shows no relation between a star's mass and its orbital energy or distance to the centre. In a system in which two-body relaxation plays no role, this state (complete or no mass segregation) persists for all time (ignoring stellar-evolutionary and tidal effects). In a system in which energy equipartition does play a role over its life time, the initially mass-segregated cluster will evolve away from the perfect initial mass segregation because stars exchange energy on a two-body relaxation timescale, while a cluster that initially is not mass segregated will evolve towards a state of statistical mass segregation because of the energy equipartition process \citep[e.g.][]{segr_aarseth,spitzer_hart_method,spitzer_hart_model,segr_obs,segr_stodol,heggie_hut,msp}. Both initial states are likely to become eventually indistinguishable.

Our work contributes to solving this problem. The type of test performed here has not been done before in that the initially perfectly mass-segregated configurations, which are consistent with the observed extremely young embedded clusters \citep[e.g.\ the Serpens South by][]{plunkett}, have not been considered in comparison with observational data. It has not been shown either on which timescale an initially perfectly mass-segregated model develops through the energy redistribution process as a result of two-body relaxation to a state that cannot be discerned at an observational level from an initially not mass-segregated state that evolves through the same process to a mass-segregated state.

We first consider the observational constraints by \citet{plunkett}, which suggest that embedded clusters that are younger than about a crossing time are perfectly mass segregated. We note that primordial mass segregation is also found by and \citet{lane_kirk_etal16}, who studied very young embedded clusters. We note in particular that the individual embedded clusters \citep[or NESTS according to][]{joncour} in the Taurus cloud are found by \citet{kirk_myers11} to be mass-segregated. Further evidence for primordial mass segregation in globular clusters (GCs) in the Milky Way was found by \citet{haghi_etal14,haghi_etal15}, and a high degree of primordial mass segregation seems to be necessary to form low-density GCs, such as Palomar 4 or 14 \citep{zon11,zon14,zon17}.
Using $N$-body simulations of idealised systems, we then compare a perfectly initially mass-segregated cluster to a not mass-segregated cluster to evaluate the timescale over which both approach the same degree of dynamical mass segregation.
The initial conditions of both systems (mass segregated or not) were set up carefully such that the systems do not evolve in a violent manner. The two methods that we use \citep{subr_model,baumgardt_segr} were tested to generate stable and reliable sets of initial conditions.
The models used here differ in the degree of primordial mass segregation but have an identical number of stars, total mass, initial mass function (IMF), and initial density profile. Therefore, we do not have to evaluate the absolute measure of mass segregation for each one of them \citep[e.g.\ through the method of the minimum spanning tree as in][]{spanning_tree}, but we may use computationally less expensive methods to compare the models to each other and evaluate a relative difference in their mass segregation.
Finally, we also compare the models to the observational data of the Orion Nebula Cluster (ONC), which is  the only currently available very young cluster in which the stellar population from the hydrogen-burning limit to O-stars has been mapped,  in order to understand whether its observed property of some mass segregation is consistent with perfect initial mass segregation or no initial mass segregation.

We note that the selected clusters used here (Serpens South and the ONC) are not two randomly chosen or extreme cases. The observed highly mass-segregated extremely young Serpens South cluster \citep[observed and reported for the first time  by][]{plunkett} is very useful concerning the question of primordial mass segregation. The ONC is not extreme, it is merely the closest very young cluster in which the full mass range of stars, from 0.1 to about $50\,\msun$, has been observed.

\section{Models}
\label{sec:models}

\begin{table}
  \centering 
  \caption{Initial parameters of the low-mass star cluster model: number of stars, mass of the cluster, half-mass radius, crossing time, and median relaxation time. The mean values for each model are given.}
  \begin{tabular}{lc}
    \hline
    model (\mseg / \texttt{\v Subr} / \mnon) & \\
    \hline
    $N$             & 52 \\
    $\mcl\ [\msun]$ & 8.0 \\
    $\rh\ [\pc]$    & 0.131 \\
    $\tcr\ [\myr]$  & 1.05 \\
    $\trel\ [\myr]$ & 0.59 \\
    \hline
  \end{tabular}
  \label{tab:params_plunkett}

\vspace*{\floatsep}

  \centering 
  \caption{Initial parameters of the star cluster models: number of stars, mass of the cluster, half-mass radius, crossing time, and median relaxation time. The mean values for each model are given.}
  \begin{tabular}{lcccc}
    \hline
    model  (\mseg / \mnon) & 1.2k   & 2.4k   & 4.7k   & 9.2k   \\
    \hline
    $N$                    & 1240   & 2404   & 4693   & 9230   \\
    $\mcl\ [\msun]$        & 659.0  & 1318.1 & 2636.0 & 5272.1 \\
    $\rh\ [\pc]$           & 0.233  & 0.254  & 0.278  & 0.305  \\
    $\tcr\ [\myr]$         & 0.274  & 0.221  & 0.179  & 0.145  \\
    $\trel\ [\myr]$        & 1.80   & 2.54   & 3.66   & 5.37   \\
    \hline
  \end{tabular}
  \label{tab:params_onc}
\end{table}

\begin{figure*}
        \centering
        \includegraphics[width=.675\linewidth]{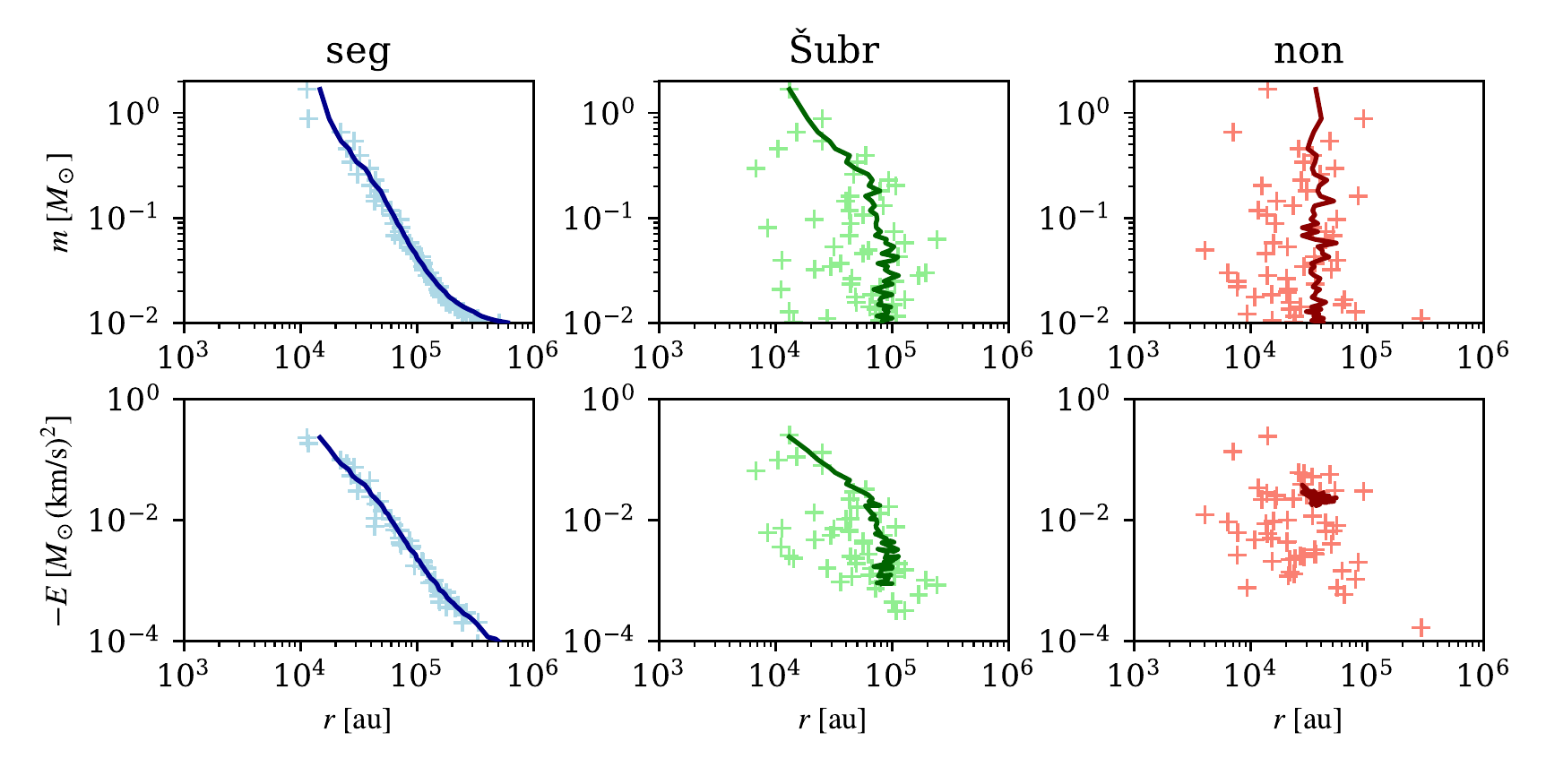}
        \caption{Initial distribution of masses (upper panels) and total energy (lower panels) with respect to the three-dimensional radial position of a star in the cluster. Three primordial mass segregations of our models containing 52 stars are shown: \mseg\ and \texttt{{\v S}ubr} are initially mass segregated with $S=1$ and $S=0.5$, respectively, and \mnon\ is without any initial mass segregation ($S=0$), see Sect.~\ref{sec:models}. The light crosses represent one realisation, and the darker lines are averages from all realisations.}
        \label{fig:mass_energy}
\end{figure*}
\begin{figure*}
        \centering
        \includegraphics[width=.9\linewidth]{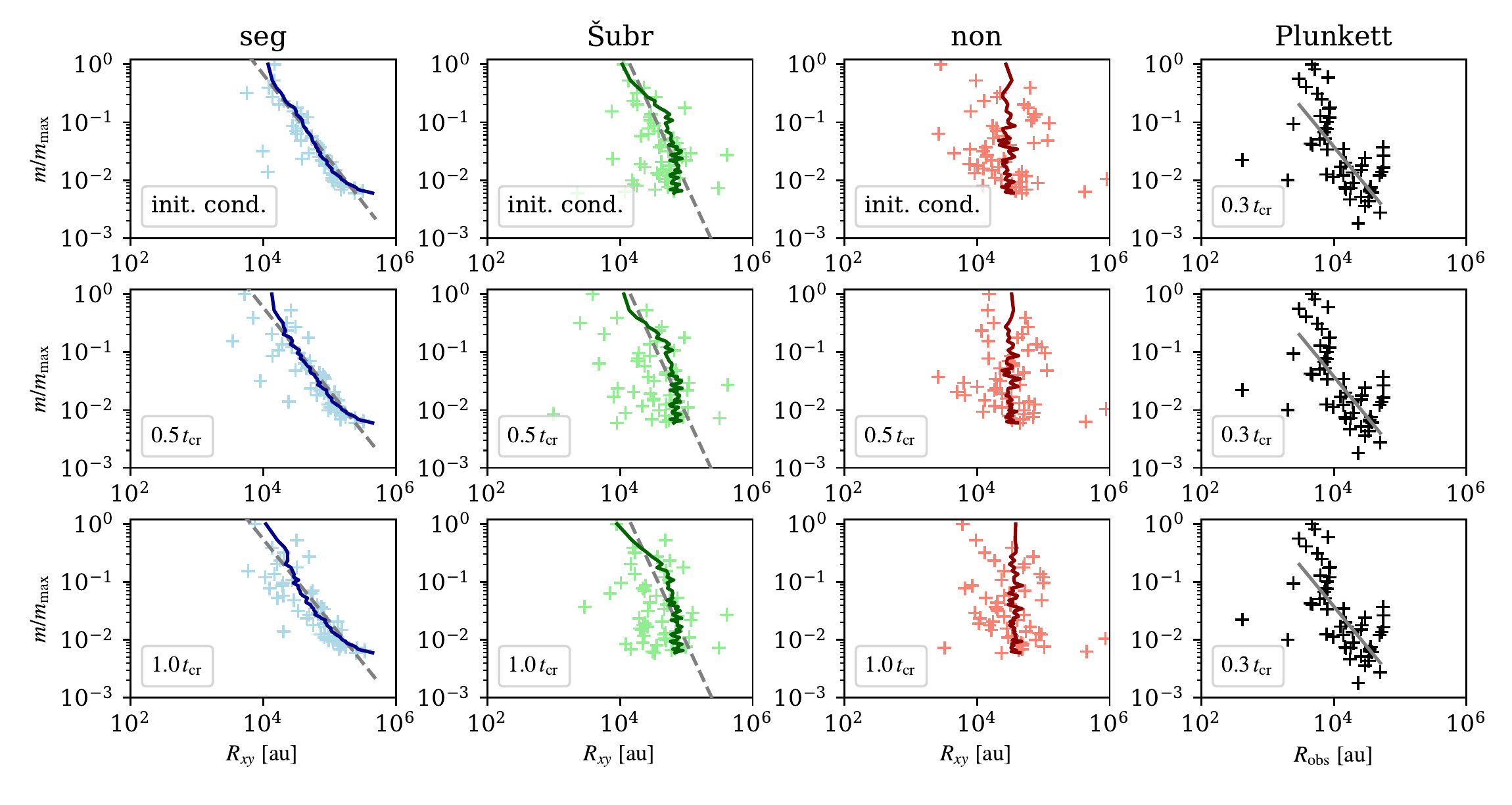}
        \caption{Three left upper panels show the initial conditions of our models containing 52 stars: \mseg\ and \texttt{{\v S}ubr} are initially mass segregated with $S=1$ and $S=0.5$, respectively, and \mnon\ is without any initial mass segregation ($S=0$), see Sect.~\ref{sec:models}. The lower panels show the models in a slightly evolved stage (at $0.5\,\tcr$ and $1.0\,\tcr$). The solid line in each plot represents the average position for a star of a given mass in 100 realisations, whereas the lighter crosses are the positions from one random realisation. The power-law fit of both \mseg\ and \texttt{{\v S}ubr} models is shown by the grey dashed line. For comparison, the right panels show the observed data and the fitted slope \citep[cf.][]{plunkett}; the same dataset is used in all rows. We note that the horizontal axis of each plot is in projection to the $xy$ plane.}
        \label{fig:plunkett_radial}
\end{figure*}

We performed numerical simulations of star clusters represented by $N$-body models (i.e.\ idealised mathematical models without gas or interstellar medium). Their initial conditions were set up using \texttt{McLuster} \citep{mcluster} as follows. The positions of individual stars were distributed according to \cite{plummer}, we used the canonical IMF \citep[][]{kroupa} with optimal sampling \citep{kroupa_etal13}. The systems are isolated, and we did not consider any additional parameters, such as a primordial binary star population or stellar evolution. The models were integrated using the collisional code \texttt{nbody6} \citep{aarseth}.

In terms of the number of stars, we created several models of young star clusters: (i) A low-mass one, containing 52 stars, see Tab.~\ref{tab:params_plunkett}, which is to be compared with the sources of the Serpens South star cluster analysed by \cite{plunkett}. This comparison will establish whether the methods for creating initial conditions are realistic. (ii) A model with 2.4k stars, which resembles a young star cluster (see Tab.~\ref{tab:params_onc}), also to be compared to the ONC. (iii) One lower mass model with 1.2k stars and two more massive models with 4.7k and 9.2k stars (see Tab.~\ref{tab:params_onc}) that serve to verify the timescale on which the initially mass-segregated clusters evolve to a state comparable to pure dynamical mass segregation.
The initial cluster size, that is, the half-mass radius ($\rh$), of all models was determined from their masses based on the birth radius-mass relation of embedded clusters by \citet{marks_kroupa}. This radius-mass relation uses the observed binary star binding energy distribution functions in very young and open clusters to derive the allowed initial densities of the clusters, which are compared to molecular clump densities and the birth densities derived independently for globular clusters. These birth densities yield the birth radius-mass relation. The total mass of our clusters, $\mcl$, is identified with the total stellar mass, $M_\mathrm{ecl}$, from \citet{marks_kroupa}. The timescales listed in Tabs.~\ref{tab:params_plunkett} and~\ref{tab:params_onc} were calculated according to \cite{aarseth}, that is,\
\begin{equation}
        \tcr \approx 2 \sqrt{2} \sqrt{\frac{2.2 \rh^3}{G \mcl}}
\end{equation}
for the crossing time, where $G \approx 4.49\times 10^{-3}\, \pc^3 \msun^{-1} \myr^{-2}$ is the gravitational constant, and
\begin{equation}
        \trel \approx \frac{0.138 N}{\ln{(0.4 N)}} \sqrt{\frac{\rh^3}{G \mcl}}
\end{equation}
for the half-mass relaxation time \citep[cf.][]{spitzer_hart_method}.

In all clusters, we analysed two extreme degrees of mass segregation, $S$, according to \citet{baumgardt_segr}: no mass segregation ($S = 0$, models are labelled as \mnon) and a fully mass segregated cluster ($S = 1$, labelled as \mseg). In the case of $S=1$, stars with the highest mass are given the lowest total energy, which places them preferentially in the centre of the cluster, while lower mass stars are given higher energies, and they therefore are on higher orbits. The cluster then looks as if it were sorted by mass and energy from the centre outwards, see the initial conditions plotted in the left panels of Fig.~\ref{fig:mass_energy}.
For $S = 0$, stars are positioned randomly and no additional sorting takes place, see the right panels of Fig.~\ref{fig:mass_energy} where the \mnon\ model average is localised in the energy space because each realisation is a random scatter. The properties of both the initially mass-segregated cluster and the initially non-segregated cluster are further discussed in Appendix \ref{ap:models}.

For a comparison of the methods we used, we also included an additional model with 52 stars using the method for mass segregation in energy by \citet{subr_model} with $S=0.5$, which is a recommended maximum value. The energy space of the \texttt{\v Subr} models (see Fig.~\ref{fig:mass_energy}) was sampled according to predefined rules as well, hence the average forms a line. We note that $S$ has a considerably different meaning in \citet{baumgardt_segr} and \cite{subr_model}.

In all models, the cluster centre is identified with the density centre provided by \texttt{nbody6}, that is,\ a numerically optimised method based on \citet{casertano_hut}.
All clusters were generated in virial equilibrium.

\section{Mass segregation}

\subsection{Initial conditions}
\label{sec:mass_plunkett}

First, we compare the small numerical model containing 52 stars with the data of the very young star cluster Serpens South observed by \citet{plunkett}, with the same number of sources. The reported age of the observed cluster is about $0.2\,\myr$ and the reported crossing time is $0.6\,\myr$.
We have integrated 100 realisations of each combination of the initial conditions (see Sect.~\ref{sec:models}) up to one crossing time of the models (see Tab.~\ref{tab:params_plunkett}). We used this star cluster only to test the method of creating a young mass-segregated cluster, that is,\ whether the model is able to reproduce the observational data to a certain degree of accuracy. We note that real star clusters tend to be formed from converging filaments \citep{andre_etal,hacar_etal1,hacar_etal2}, thus are neither spherically symmetric nor Plummer models, although the filament cross-sections have a Plummer-like profile \citep{andre_etal}, and that the choice of IMF could play a role as well. The evolution on a crossing timescale is sufficient to see also whether the methods we used can create reasonably stable initial conditions.

In Fig.~\ref{fig:plunkett_radial} we indicate with crosses the projected radial coordinate of each star versus the ratio of its mass to the most massive star or source in the cluster for one realisation. As the IMF is optimally sampled, the masses of stars are the same in all 100 realisations. Thus, we also show the average position of a star of a given mass with a solid line. Because the models are spherically symmetric, there is no preferred plane of projection. Thus, we took the $xy$ plane of coordinates of the model. First, we may see that the initial conditions are reasonably stable during the first crossing time (the averaged positions stayed almost the same).
Although we tested only idealised models, we see after visually comparing the plots in Fig.~\ref{fig:plunkett_radial}  that our segregated models (two left panels) do represent the observations (right panel) better than the non-segregated model (middle right panel) at $t=0$, $0.5$ and even $1.0\,\tcr$. This is also confirmed with a quantitative comparison of the power-law functions fitted to the data. The observed data have a power-law index of $-1.4 \pm 0.2$ \citep[][verified here]{plunkett}; in all time frames, the mean slope in the log-log plot of the \mseg\ model is $-1.44 \pm 0.13$, and $-2.46 \pm 0.14$ for the \texttt{\v Subr} model; the \mnon\ model has virtually a vertical slope, which makes the fitting ambiguous and the model not compatible at all with the observed data.

Hence, we conclude that the observational data of \citet{plunkett} show a most impressive degree of mass segregation that is comparable to the perfectly mass-segregated models ($S=1$).
We note that independent authors also reported that very young embedded clusters are mass segregated (see Sect.~\ref{sec:intro}).

\subsection{Evolution}
\label{sec:mass_seg}

\begin{figure}
        \centering
        \includegraphics[width=\linewidth]{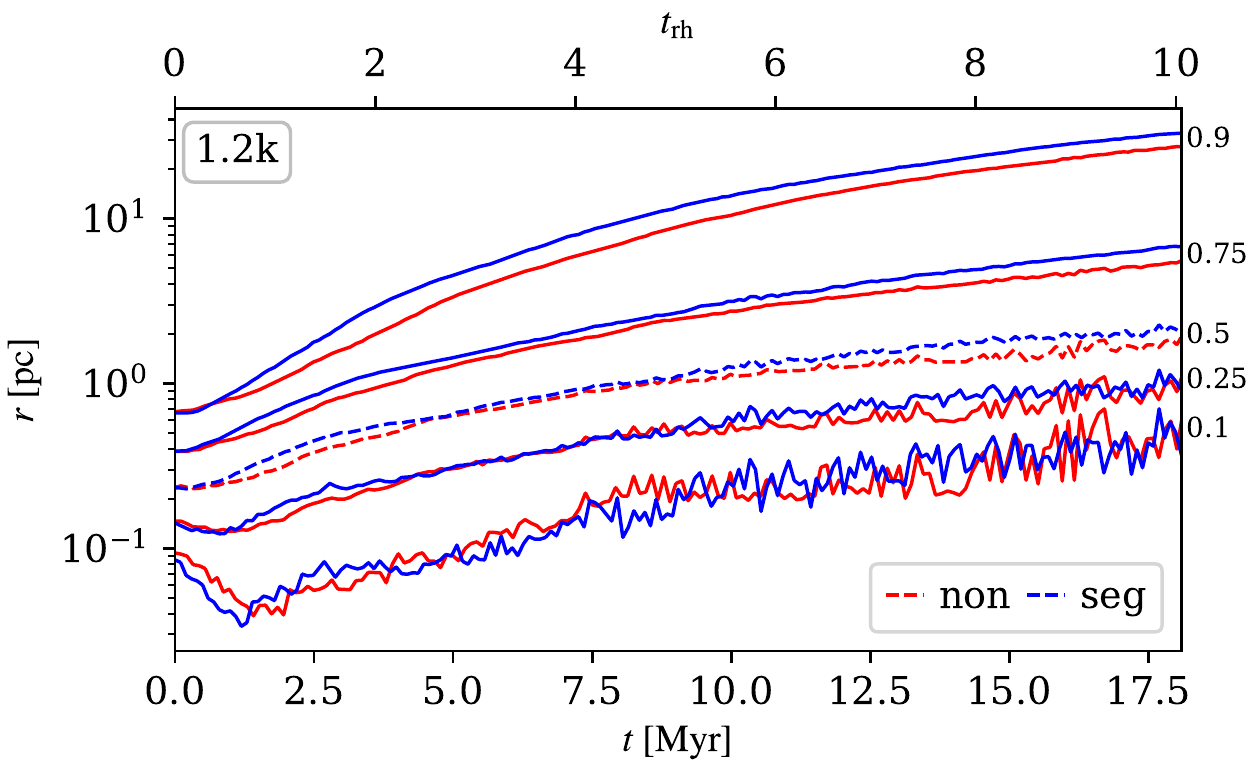}\\
        \includegraphics[width=\linewidth]{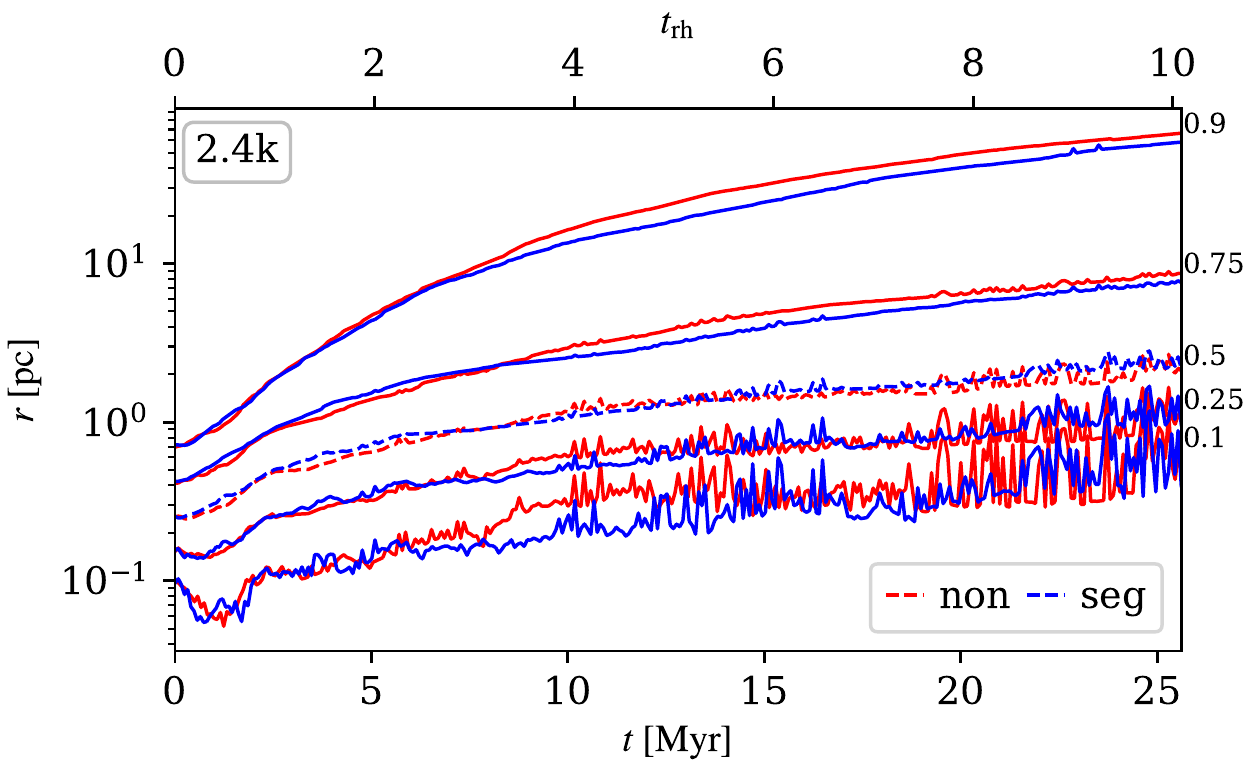}\\
        \includegraphics[width=\linewidth]{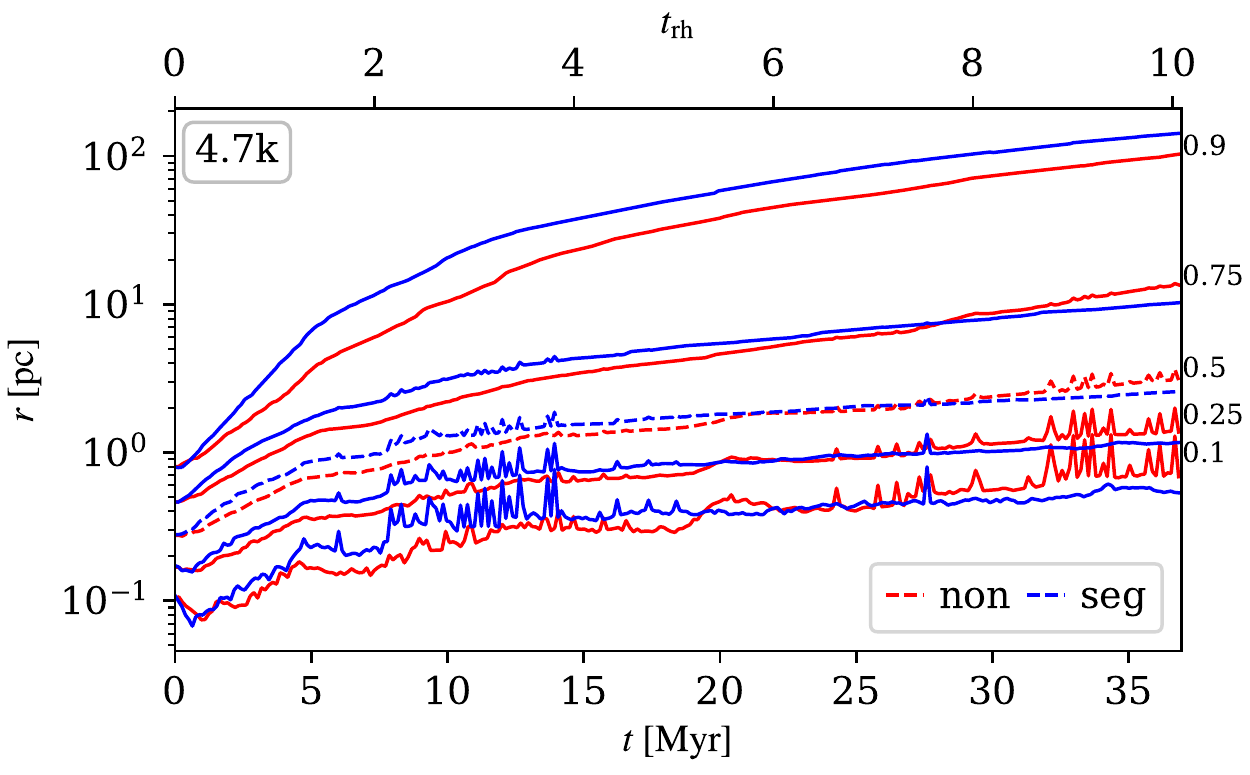}\\
        \includegraphics[width=\linewidth]{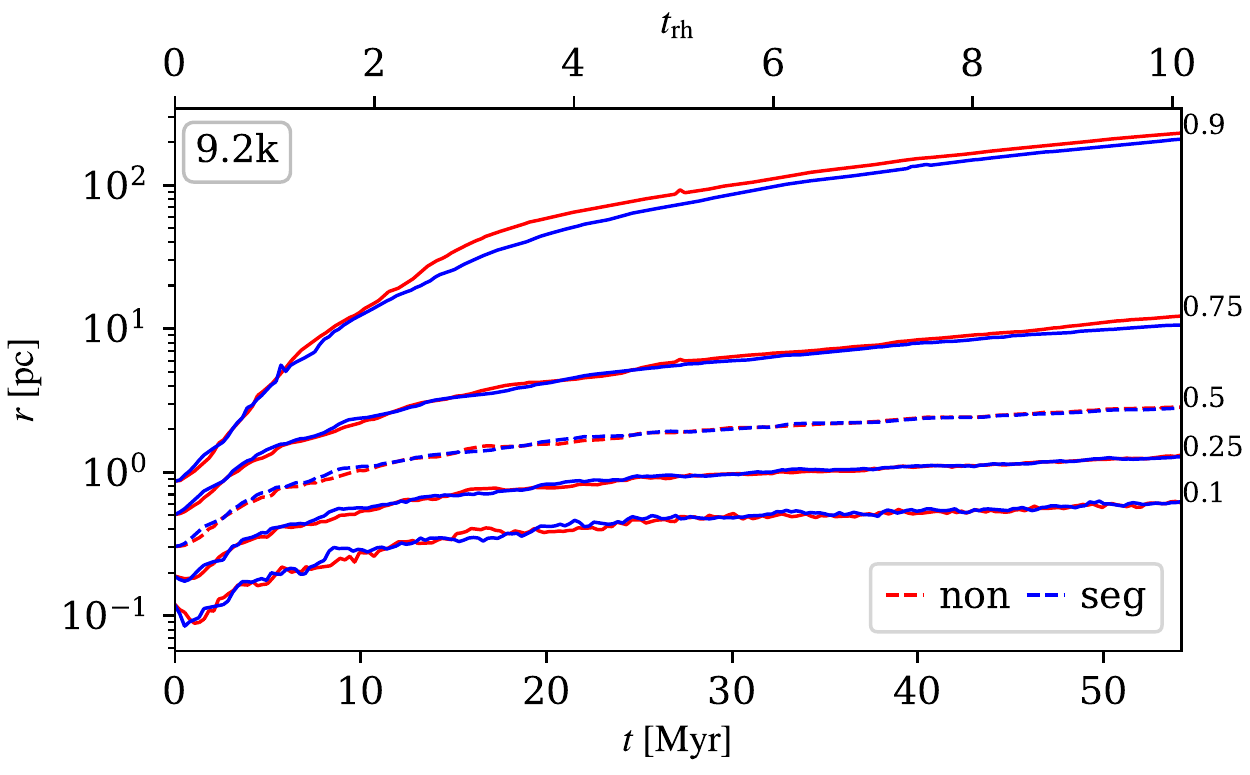}
        \caption{Lagrangian radii of both segregated (blue) and non-segregated (red) models with 1.2k, 2.4k, 4.7k, and 9.2k stars (panels from top to bottom), averaged over all realisations of each model. The time axis reaches ${\approx}10\,\trel$, the corresponding mass fractions (10, 25, 50, 75, and 90\,\% from bottom to top) are on the right-hand side of each panel, and the half-mass radius is plotted with a dashed line.}
        \label{fig:lagr}
\end{figure}

In the case of the star clusters with 1.2k stars, we have 20 realisations, for 2.4k, 4.7k, and 9.2k stars, we have integrated ten realisations of each combination of the initial conditions. The initial parameters are listed in Tab.~\ref{tab:params_onc}.
The whole cluster evolution can be visualised using the Lagrangian radii; the average over all realisations is plotted in Fig.~\ref{fig:lagr} for each model.
The mass-segregated models undergo core collapse slightly earlier than the non-segregated models, the latter lagging behind, which is due to the need of dynamical mass segregation to first gather massive stars in the core.
This is most visible in the smallest model with 1.2k stars, where two-body encounters between massive stars have a higher impact on the whole system than in a larger model. During the core collapse, initial differences in the distribution of the most massive stars ($>5\msun$) of both \mseg\ and \mnon\ models are smeared out. Thus, after the core collapse (at about 2.5 to 3\,Myr), both \mseg\ and \mnon\ Lagrangian radii appear to be the same in all models (except for random fluctuations, especially in the core region). Differences in the distribution of stars with lower masses still remain at this point, however. We further investigate this in each model in Appendix~\ref{ap:segr}.

\begin{figure*}[!htbp]
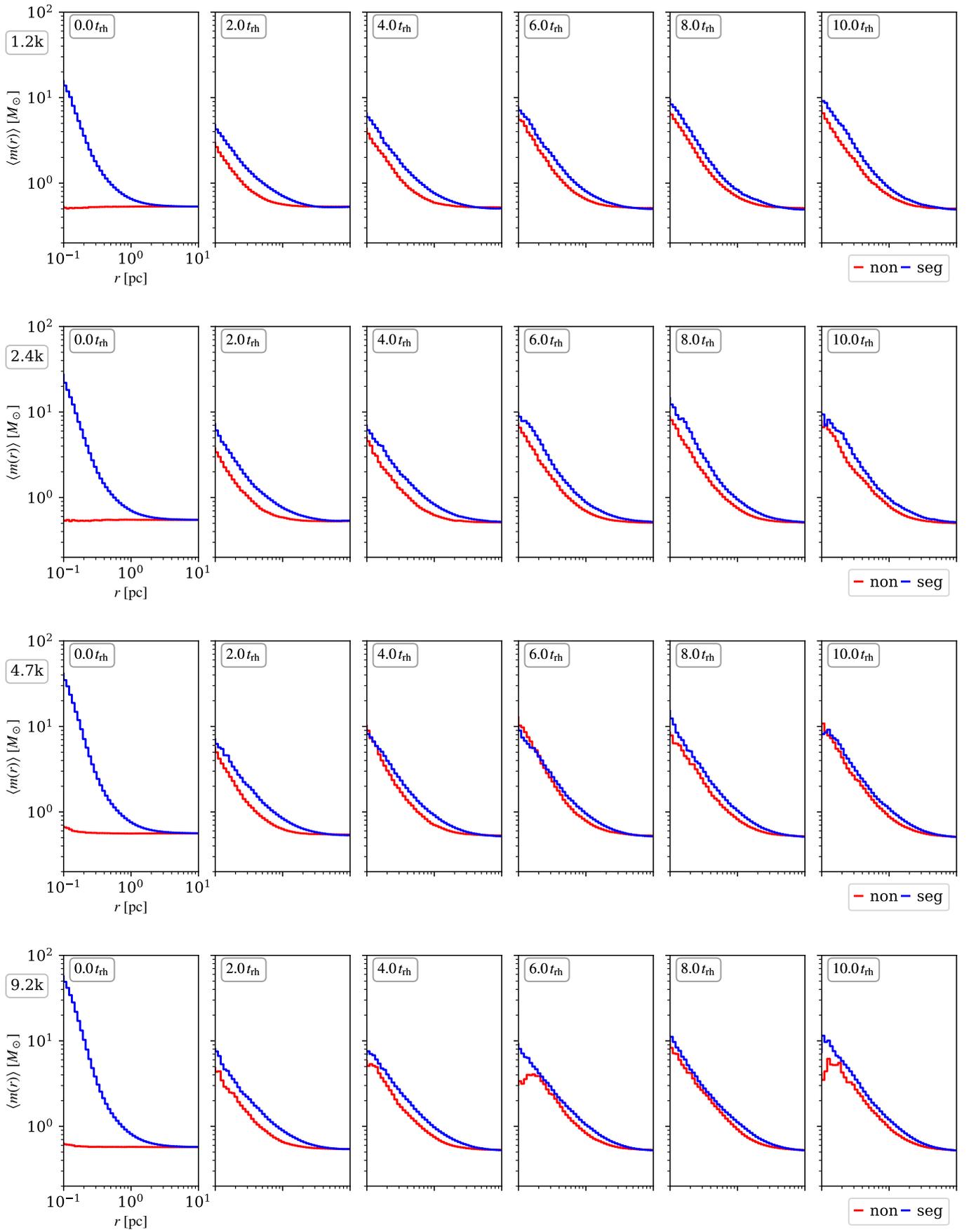

        \centering
        \includegraphics[width=\linewidth]{{{m659sun_mean_mass_bin_10.0_p0_0-18}}}\\
        \hspace{1pt}
        \includegraphics[width=\linewidth]{{{m1318sun_mean_mass_bin_10.0_p0_0-25}}}\\
        \hspace{1pt}
        \includegraphics[width=\linewidth]{{{m2636sun_mean_mass_bin_10.0_p0_0-36}}}\\
        \hspace{1pt}
        \includegraphics[width=\linewidth]{{{m5272sun_mean_mass_bin_10.0_p0_0-54}}}
        \caption{Mean mass contained in a sphere of a given radius, see Eq. \eqref{eq:mean_mass}, of the models with 1.2k, 2.4k, 4.7k, and 9.2k stars (panels from top to bottom), averaged over all realisations of each model. Both initially mass-segregated (blue) and not mass-segregated clusters (red) of each model are plotted. Individual plots are separated by $2.0\,\trel$.}
        \label{fig:mean_mass}
\end{figure*}

\begin{figure*}[!htb]
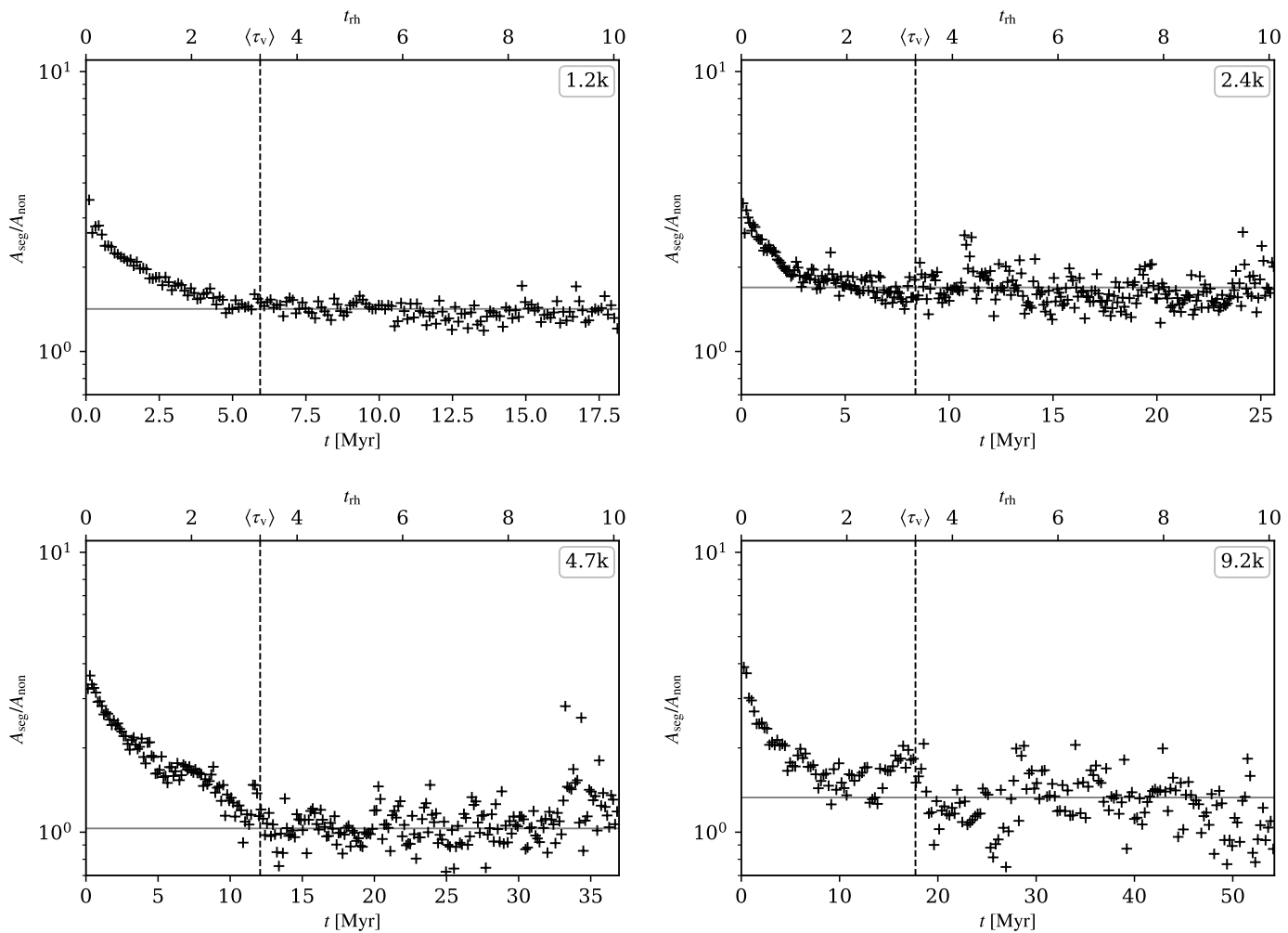

        \centering
        \includegraphics[width=.495\linewidth]{{{m659sun_seg_par_bin_0.1-10.0}}}\hfill
        \includegraphics[width=.495\linewidth]{{{m1318sun_seg_par_bin_0.1-10.0}}}\\
        \includegraphics[width=.495\linewidth]{{{m2636sun_seg_par_bin_0.1-10.0}}}\hfill
        \includegraphics[width=.495\linewidth]{{{m5272sun_seg_par_bin_0.1-10.0}}}
        \caption{Evolution of the ratio given by Eqs.~\eqref{eq:mean_mass} \&~\eqref{eq:area} in time. The dashed line and the value $\langle\tauv\rangle$ (see Eq.~\eqref{eq:tauv}) represent the mean time when the slope of the data points turned zero. The slope itself is plotted by a grey line.}
        \label{fig:tau_v}
\end{figure*}

The difference between a collisional system that is and is not initially mass segregated gradually vanishes as the systems evolve. This is better illustrated using the radial mass distribution. In Fig.~\ref{fig:mean_mass} we plot the mean mass contained in concentric spheres of radii (from above the core region to several $\rh$) for each model from Tab.~\ref{tab:params_onc}. Here, we merge all realisations of a particular model and divide the stars into logarithmically equidistant bins, that is,\ for a $k$-th bin at radius $r_k$, the mean mass is
\begin{equation}
        \label{eq:mean_mass}
        \langle m(r_k) \rangle = \frac{\sum_{i = 1}^k{m_i}}{\sum_{i = 1}^k{n_i}} \,,
\end{equation}
where $n_i$ and $m_i$ are the number of stars in an $i$-th bin and their total mass, respectively.
Initially, we observe a clear difference because both models are generated with a different degree of mass segregation. Then, both cluster profiles approach each other due to energy redistribution; the non-segregated model gradually develops mass segregation, while the other model looses its primordial mass segregation. Visually, the distributions start to look similar at $\approx 2\,\trel$. If we omit oscillations of the core region (i.e.\ close to 0.1\,pc), the following evolution of both \mseg\ and \mnon\ models looks the same for several relaxation times (in the case of the more populous clusters, it is almost perfectly synchronous).

The fluctuations of the mean mass at low radii are higher for the most massive model, as can be seen in the lower panels of Fig.~\ref{fig:mean_mass}. In the more massive models, the \mseg\ and \mnon\ curves of mean mass almost coincide and frequently cross, whereas in the lower mass models, they keep a fixed distance from each other.

In order to quantify the difference between the \mseg\ and \mnon\ models, we compare the histograms given by Eq.~\eqref{eq:mean_mass} and presented in Fig.~\ref{fig:mean_mass}. The most important data that show mass segregation are at the smallest radii (i.e.\ near the cluster core). Because the histograms have equidistant bins in log-scale, large differences at small radii could be overshadowed by small differences at larger radii if we just summed their heights. Therefore, we weight each bin by its inverse width instead, which ensures that the bins at smaller radii become more important than the bins at larger radii, which gives us the parameter
\begin{equation}
        \label{eq:area}
        A = \sum_{k = 1}^{\nbin}{\frac{\langle m(r_k) \rangle}{\Delta r_k}} \,,
\end{equation}
where $\Delta r_k$ is the width of the $k$-th bin (with logarithmically equidistant bins) and $\nbin$ is the total number of bins. In particular, $r_1 = 0.1\,\pc$, $r_{\nbin} = 10\,\pc$ and $\nbin = 50$ for all models, as plotted in Fig.~\ref{fig:mean_mass}. A further discussion of the choice of binning and weights follows in Appendix~\ref{ap:bins}.
At each time step, we evaluate the ratio $A_\mseg / A_\mnon$ of the initially mass-segregated and not mass-segregated clusters. This ratio, which is plotted for each model in Fig.~\ref{fig:tau_v}, decreases initially and then settles at an almost constant value close to $A_\mseg / A_\mnon \approx 1$. Therefore, the initial difference of mass segregation almost vanishes (we denote this time $\tauv$).
To derive the exact value of $\tauv$ in each model, we fitted the time after which the data points correspond to a horizontal line (given the fluctuations of the data, the uncertainty of the fitted slope was set to ${\lesssim}10^{-3}$ in the lower mass models and ${\lesssim}10^{-2}$ in the higher mass models; these uncertainties correspond to the scatter that is evident in Fig.~\ref{fig:tau_v}). The values that we obtained for each model are similar in units of the initial median relaxation times. Thus, we present an empirical estimate on the time when the primordial mass segregation vanishes in the range
\begin{equation}
        \label{eq:tauv}
        3\,\trel < \tauv < 3.5\,\trel \,,
\end{equation}
with the mean value around $\langle\tauv\rangle \approx 3.3\,\trel$ which is also plotted for reference in Fig.~\ref{fig:tau_v}. We note that the value of $\tauv$ seems to be identical in all models.

The shallower potential well and lower escape velocity of the lower mass models implies that these clusters are more effective in ejecting stars from the central region as the individual two-body encounters are stronger. This is the case especially in the primordially mass-segregated clusters where the concentration of high-mass stars in the core is higher. On the other hand, in the higher-mass clusters, a collective effect of many two-body interactions is often needed to eject a star \citep[cf.][]{oh_etal15}.
Consequently, the initially mass-segregated low-mass clusters inflate more than the non-segregated low-mass clusters or higher mass clusters. The mean mass will, therefore, decrease at smaller radii and increase at higher radii, and the values of $\langle m(r_k) \rangle$ of an initially mass-segregated cluster will stay above those of an initially non-segregated cluster.
This is also visible in the half-mass radii in Fig.~\ref{fig:lagr}, where the half-mass radius of the mass-segregated 1.2k cluster is above the non-segregated one even after $10\,\trel$. Proceeding towards more massive models, this difference becomes less visible or disappears completely. Thus, the ratio of $A_\mseg / A_\mnon$ settles at a higher value than $1$ in the lower mass models.

\section{Orion Nebula Cluster}
\label{sec:data}

The question whether embedded star clusters form fully mass-segregated is approached here using the best observed (and nearest) case that formed stars from $0.1\,\msun$ to the O-star regime: the ONC.
It has been  established for the present-day ONC that it is mass segregated in the range above $5\,\msun$ by \citet{hillenbrand}, which was  later confirmed by \citet{spanning_tree}, who used the minimum spanning tree method on the data from \citet{hillenbrand_hartmann}. In this section, however, we focus on the possibility of a complete primordial mass segregation, as discussed in the preceding section.


\subsection{Datasets}

\setlength\tabcolsep{3.75pt}
\begin{table*}
  \centering 
  \caption{Data of the ONC used in this work. Column names and several lines are shown to demonstrate the data structure. The complete dataset is available in a computer-readable format.}
  \begin{tabular}{ccccccccccccccc}
    \hline
    RA $[\adeg]$ & DEC $[\adeg]$ & [H97b] & [HC   & [MLLA] & \multicolumn{3}{c}{[FDM2003]} & [DRH & [COUP] & [PMF &
    $M_\mathrm{H97b}$ & $M_\mathrm{FDM2003}$ & $M_\mathrm{B98}$ & $M_\mathrm{DM98}$ \\
                   &     &        & 2000] &        & Opt & X & X$_2$               & 2012]&        & 2008]&&&&\\
    \hline
                83.617\dots & $-5.444\dots$ & 1   &     &     & 22  &     & &      & & & 0.11 & 0.15 &       &        \\
                83.618\dots & $-5.416\dots$ & 2   &     &     & 23  &   8 & &      & & & 0.67 & 1.20 &       &        \\
                83.618\dots & $-5.649\dots$ &     &     &     &     &     & & 1267 & & &      &      & 0.038 & 0.055  \\
                83.620\dots & $-5.244\dots$ &     &     &     &     &     & &  784 & & &      &      &       & 0.700  \\
                83.776\dots & $-5.519\dots$ & 294 &     &     & 228 &     & &      & & & 0.15 & 0.21 &       &        \\
                83.776\dots & $-5.411\dots$ & 291 &  99 & 161 & 230 & 136 & &      & & & 0.27 & 0.43 &       &        \\
                83.803\dots & $-5.345\dots$ & 391 & 699 & 916 & 293 & 214 & 222 &  & & & 0.29 & 0.43 &       &        \\
                83.857\dots & $-5.493\dots$ & 789 &     &     &     &     & &      & & & 0.20 &      &       &        \\
                83.709\dots & $-5.442\dots$ &     &     &     &     &     & &   & 53 & &      &      &       &        \\
                83.807\dots & $-5.359\dots$ &     & 602 & 797 &     &     & &  & 570 & 34 &   &      &       &        \\
                \dots &&&&&&&&&&&&&&\\
    \hline
    \multicolumn{15}{l}{\footnotesize [H97b] is the Simbad identifier of the data from \cite{hillenbrand} with multiplicity index}\\
    \multicolumn{15}{l}{\footnotesize [HC2000] is the Simbad identifier of the data from \cite{hillenbrand_carpenter}}\\
    \multicolumn{15}{l}{\footnotesize [MLLA] is the Simbad identifier of the data from \cite{muench}}\\
    \multicolumn{15}{l}{\footnotesize [FDM2003] is the Simbad identifier of the data from \cite{flaccomio_a,flaccomio_b}, where Opt is for the optical, X is for the X-ray,}\\
    \multicolumn{15}{l}{\footnotesize \hspace{43pt} and X$_2$ is the identifier of an additional X-ray source that was also cross-matched to the optical data}\\
    \multicolumn{15}{l}{\footnotesize [DRH2012] is the Simbad identifier of the data from \cite{dario}}\\
    \multicolumn{15}{l}{\footnotesize [COUP] is the Simbad identifier of the data from \cite{COUP}}\\
    \multicolumn{15}{l}{\footnotesize [PMF2008] is the Simbad identifier of the data from \cite{prisinzano}}\\
    \multicolumn{15}{l}{\footnotesize $M_\mathrm{H97b}$ is the mass given in \cite{hillenbrand}}\\
    \multicolumn{15}{l}{\footnotesize $M_\mathrm{FDM2003}$ is the mass given in \cite{flaccomio_a,flaccomio_b}}\\
    \multicolumn{15}{l}{\footnotesize $M_\mathrm{B98}$ is the mass given in \cite{dario} from \cite{BCAH98}}\\
    \multicolumn{15}{l}{\footnotesize $M_\mathrm{DM98}$ is the mass given in \cite{dario} from \cite{DM98}}\\
  \end{tabular}
  \label{tab:data}
\end{table*}
\setlength\tabcolsep{6pt}

For the purpose of comparing our evolutionary models with the observational data, we used a publicly available database\footnote{Downloaded from Vizier and Simbad:\\ \url{http://vizier.u-strasbg.fr/viz-bin/VizieR}\\ \url{http://simbad.u-strasbg.fr/simbad}.}. The sources of data are listed in Tab.~\ref{tab:data}. The structure of the ONC is complex: stars are surrounded and also partially covered by the interstellar medium (part of the Orion molecular cloud \citep{hillenbrand, hillenbrand_hartmann}), which results in varying extinction in different wavelengths \citep{scandariato_et_al}. Therefore, a combination of optical photometry and spectroscopy with X-ray and \hbox{(near-)IR} observations is used for more reliable results. We are also aware that our models are purely mathematical representations of the real ONC, nevertheless, we can use them to understand basic properties of dynamical evolution in the observed cluster.

In order to construct the most complete dataset of the ONC, we combined several existing catalogues that are listed in Tab.~\ref{tab:data}. The main component is the optical dataset of \citet{hillenbrand}, which offers a sample of 1576 stars in the Orion Nebula region. Masses are determined for 929 stars using the evolutionary models of \cite{DM94}, \cite{swenson_etal}, and \cite{ezer_cameron}. The membership probability is based on proper motion \citep[e.g.][]{jones_walker}, and we excluded those labelled as probable non-members. Stars that were farther than $18.29'$ from $\theta_1$ Ori were also excluded from the dataset \citep[cf.][]{hillenbrand}; this left 1176 sources in total.

We also used all 1059 sources from \cite{muench}, who provided the most complete dataset of the ONC in the IR ($K$-band), that is,\ in the range from O and B stars down to near the deuterium-burning limit. The sources are already cross-identified with \cite{hillenbrand} and \cite{hillenbrand_carpenter} observations. Stars that were labelled non-members by \cite{hillenbrand} but were labelled members by \cite{muench} are included, using coordinates from the latter (this is mostly the case of sources with IDs from [H97b]\,3000 to [H97b]\,5999).

\citet{flaccomio_a,flaccomio_b} identified 742 objects in the X-ray band and 696 in the optical wavelengths. The masses of all stars from the optical sample were estimated, and the stars are likely ONC members. However, neither the X-ray nor the optical sources were cross-identified with earlier catalogues. Therefore, we used the CDS X-Match Service\footnote{\url{http://cdsxmatch.u-strasbg.fr/}} with the fixed distance criterion ($d<3''$) to cross-identify these two datasets between each other and also with \cite{hillenbrand}. Sources from the optical measurements that were identified with multiple counterparts were further distinguished based on their masses. In some cases, the optical data were identified with two X-ray sources (i.e.\ about ten cases); both are assigned to this one source (the entries are marked in column X$_2$ in Tab.~\ref{tab:data}). Again, stars that were in the optical sample of \citet{flaccomio_a,flaccomio_b} but were marked non-members by \cite{hillenbrand} were returned to the dataset.

A majority of sources from the most complete X-ray survey of the Orion Nebula region, that is,\ COUP\footnote{Chandra Orion Ultradeep Project} \citep{COUP}, have been matched with previous optical \citep[e.g.][]{hillenbrand} near-IR or IR catalogues \citep[e.g.][]{muench,2MASS}. However, 16 new sources from COUP without counterparts were identified as ONC members by \cite{getman}.

The young stellar objects were observed for instance by \cite{prisinzano}, who identified 45 sources in the ONC. Thirteen of these have no optical, IR, or X-ray counterparts. These sources were also added to our dataset.

The IR and optical dataset from \cite{dario09} has also previously been matched with previous catalogues, therefore we did not include these sources in our ONC dataset. However, we included brown dwarfs and pre-main-sequence stars from \cite{dario} and used their cross-identification with \cite{dario09} in order to remove duplicates. Here, we also distinguished between probable members of the ONC and the foreground/background contamination based on reddening \citep{alves_bouy}. This gave us additional 163 sources that have not been observed in the optical or IR wavelengths before. The masses of all of the sources from \cite{dario} were deduced from the evolutionary tracks of \cite{DM98} and \cite{BCAH98}; a vast majority lie below $1\,\msun$.

In total, our dataset of the ONC contains 2430 sources, see Tab.~\ref{tab:data}. Out of these, 995 have estimated masses. Nevertheless, we assume that all high-mass stars (i.e.\ $m \geq 5\,\msun$) and even a majority of stars in the range from 1 to $5\,\msun$ have known masses \citep[cf.][]{hillenbrand}. Stars with an unknown mass are therefore assumed to be less massive than $5\,\msun$. This is a reasonable assumption given that $m > 1\,\msun$ stars are the brightest and thus most conspicuous cluster members. In case of multiple mass estimates for one source, we took their mean value. Because the masses have very similar values and they never cross the boundary of $5\,\msun$ (on which we focus here) in between different models, this is a valid assumption. Out of the most massive stars, we have 11 in total with $m>5\,\msun$ (and 105 with $1<m\leq 5\,\msun$).

Although there is evidence for stars in the ONC that are more than 5\,Myr old \citep{huff_stahler} or even up to 12\,Myr old \citep{warren}, they were most likely captured by the cluster potential \citep{pflamm_old_stars}.
The age of the ONC, which is estimated from the time of the most active star formation (about 1 to 2\,Myr ago), is about 2.5\,Myr \citep{hillenbrand,palla_stahler}.
Nevertheless, we see that the ONC did not form in one burst \citep{palla_etal}. It has even been shown that the ONC is likely to contain three populations of stars that are younger than 3\,Myr \citep{beccari_pop,kroupa_onc,long_etal18}.

To convert angular dimensions into a physical size scale, we need to know the distance to the ONC. We adopted the value $(414 \pm 7)\,\pc$ of \citet{menten}, who measured trigonometric parallaxes of stars in the ONC. This value is consistent with previous works \citep[cf. references in][]{menten} and also with the most recent astrometric observations of stars in the Orion~A molecular cloud from the \emph{Gaia}\footnote{Global astrometric interferometer for astrophysics} mission \citep{gaia_dr2}, where \citet{gaia_orion} obtained a distance to the ONC between 400\,pc and 410\,pc. For $414\,\pc$, the conversion between angular and proper separation reads $1\,\amin\,{=}\,0.1204\,\pc$. This means that the ONC has a radius of approximately 2.5\,pc (this corresponds to a Lagrangian radius of ${\approx}90\,\%$).

The total number of sources in the ONC is comparable to our model with 2.4k stars. According to the results from Sect.~\ref{sec:mass_seg} (ad Fig.~\ref{fig:tau_v}), the expectation therefore is that the ONC has not yet reached an age such that a full primordial mass segregation becomes indistinguishable from an initially not segregated condition, that is,\ the time $\langle\tauv\rangle \approx 8.4\,\myr$. If the ONC did form fully mass segregated, then it ought to still show evidence for this.

\subsection{Mass segregation}

We represent the ONC using our models with 2.4k stars. In order to compare the initially mass-segregated and non-segregated model to the observational data, we plot the averaged normalised cumulative distributions of stars in Fig.~\ref{fig:onc_radial}. Each model is plotted at three different times: the initial conditions at 0\,Myr (for reference), at 1.5\,Myr, and at 2.5\,Myr, which is assumed to be the age of the ONC.
The discussion in Appendix~\ref{ap:segr} shows that both of these extreme models evolve after 2.5\,Myr (${\approx}0.5\,\trel$) to a stage where their high-mass stars ($m \geq 5\,\msun$) are distributed almost identically. This holds even in projection, where the level of mass segregation is always lower than in 3D.
The radial profiles of their low-mass stars ($m < 5\,\msun$) also seem to tend to the observed data as time proceeds, but at a much slower rate, and the shapes of the cumulative distributions are different. It is expected that less massive stars mix more slowly and thus lose their memory of the primordial mass segregation later than the more massive stars.

We evaluated the resemblance of the radial profiles of our models and the observational data using a two-sample Kol\-mo\-go\-rov--Smir\-nov (KS) test\footnote{The algorithm is implemented in the \texttt{python} library \texttt{scipy.stats} as \texttt{ks\_2samp()}.}. From the results plotted in Fig.~\ref{fig:onc_ks} from 0 to 5\,Myr, it is evident that the observed data are not consistent with either of our models. Although the $p$-value of high-mass stars of both \mseg\ and \mnon\ models increases above 0.8 after about 0.75\,Myr and then stays very high, the $p$-value for lower mass stars is of the order of $10^{-4}$ and therefore the null hypothesis (\mseg\ or \mnon) is not compatible with the data.

Several issues need to be discussed when we compare the models with observations, however. (i) The ONC is partially covered by a molecular cloud with extinction $A_V > 5\,\magn$ and in some areas even up to $A_V \approx 11\,\magn$ \cite[cf.][]{scandariato_et_al}. This layer of opaque material extends from 0.5 to about 2\,pc in approximately one quadrant of the nebula. The optical measurements in these parts of the cluster, for example, will therefore be always incomplete. This feature is visible even in the radial profile of low-mass stars (the dashed black line in Fig.~\ref{fig:onc_radial}), where the cumulative distribution suddenly changes slope above 0.5\,pc. (ii) As documented by \cite{hillenbrand_hartmann}, the ONC is not spherically symmetric in the region beyond 0.5\,pc from the centre. Therefore, it cannot be straightforwardly compared with a spherically symmetric model. (iii) Although we cross-correlated more than 20 years of observational data from the X-ray, optical, and \hbox{(near-)IR} wavelengths and compiled the possibly most complete dataset of the ONC so far, incompleteness in the form of observational bias may still play a role (cf.\ the data references). However, in comparison to the previous issues, this problem is minor.

\begin{figure}
        \centering
        \includegraphics{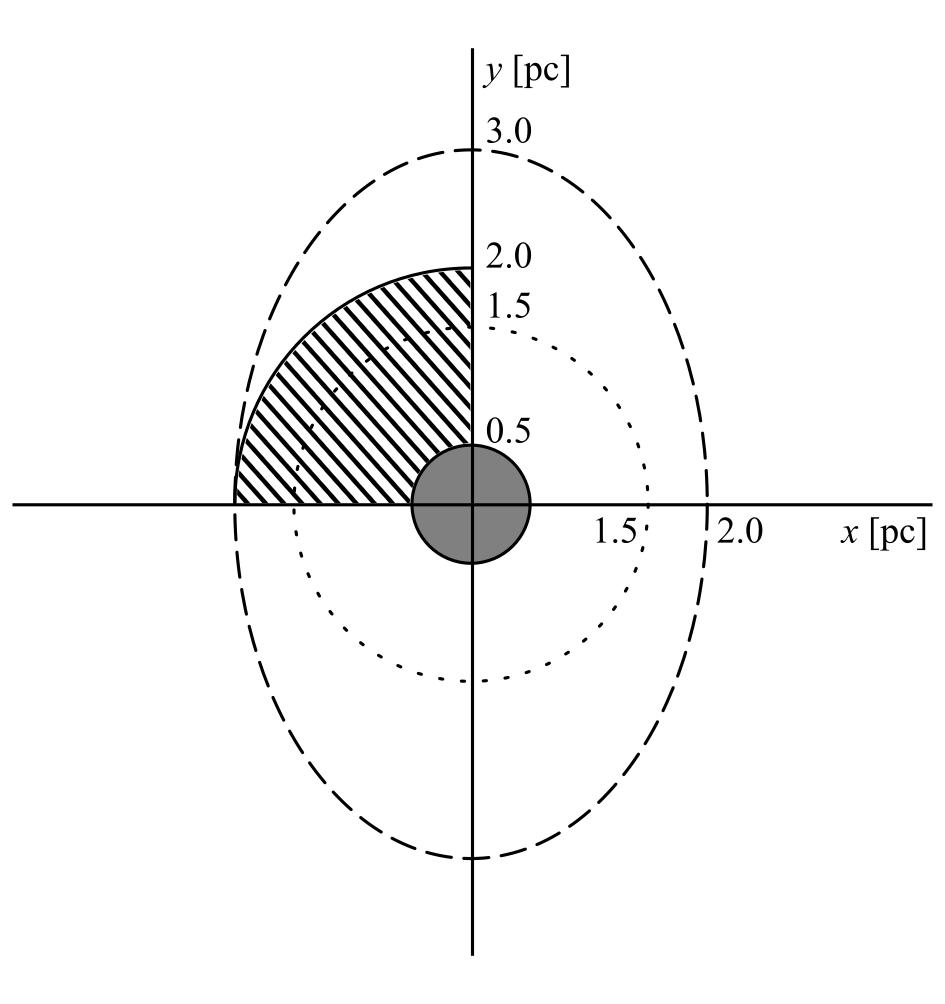}
        \caption{Schematic plot of the modifications made to the models in order to compare them to the ONC. The hatched area has artificial extinction of low-mass stars (i.e.\ stars with $m<5\,\msun$ are removed), the dashed ellipse demonstrates the scaling of low-mass stars along the $y$-axis by a factor of 1.5, the dotted circle illustrates the isotropic scaling of all high-mass stars, and the grey circle covers the central region where low-mass stars are left without scaling.}
        \label{fig:model_ext}
\end{figure}

In order to quantify the role of extinction and the non-spherical nature of the ONC, we evaluated these effects in two steps:
(i) We introduced an artificial extinction in our projected data. All low-mass stars ($m<5\,\msun$) in the area from 0.5 to 2\,pc, where $x<0$ and $y>0$, were removed, see the hatched area in Fig.~\ref{fig:model_ext}. The new radial profiles that include this extinction are plotted in the top panels of Fig.~\ref{fig:onc_radial_extscl}. When we compare them to the former radial profiles (Fig.~\ref{fig:onc_radial}), we see that the gap between the black and blue dashed curves is smaller. However, the KS test still shows no resemblance with the observational data, see Fig.~\ref{fig:onc_ks_extscl}. Because this extinction removes only up to 10\,\% of stars from the model, it is not the main responsible factor that would ensure that our model and the data are incompatible.

(ii) In addition to extinction, we also scaled the positions of low-mass stars ($m<5\,\msun$) in the model beyond a radius of 0.5\,pc along the $y$-axis by a factor of $s_y = 1.5$ in order to achieve the same elongated shape as documented by \citet[][cf.\ their Fig.~3]{hillenbrand_hartmann}. The scaling is illustrated in Fig.~\ref{fig:model_ext}.
Given that this scaling is dynamically not self-consistent, the conclusions based on this should be seen only as suggestive.
If the length scales by $s_r$, the time must be scaled by $s_t = s_r^{3/2}$. Because we scale the projected data in only one direction, we define
\begin{equation}
        \label{eq:s_r}
        s_r = \frac{\sqrt{2}}{2} \sqrt{1 + s_y^2} \,,
\end{equation}
which gives, in particular, $s_r = 1.275$ and $s_t = 1.439$.
By doing so, we approximated the correct timescale for the low-mass stars in the outer cluster ($>0.5\,\pc$), but are slowed down their evolution of the central region ($<0.5\,\pc$).
We scaled the $x$ and $y$ coordinates of high-mass stars equally in all directions by the factor $s_r$ from Eq.~\eqref{eq:s_r} because they are segregated towards the central region and no asymmetry in their distribution has been found \citep[e.g.][]{hillenbrand,hillenbrand_hartmann}. In Fig.~\ref{fig:onc_radial_extscl} (middle and bottom panels), we also plot a comparison of the radial profiles with an artificial extinction and elongated shape at three times (in the panels with scaling, two time stamps are given: first, the real time of the simulation, which also corresponds to the central region of low-mass stars, then the time scaled by $s_t$ , which is relevant for all high- and low-mass stars above 0.5\,pc). Using the KS test, we found that scaling of our models (\mseg\ and \mnon) alone is not able to describe the observational data of low-mass stars either (see the results in Fig.~\ref{fig:onc_ks_extscl} and the radial profiles in Fig.~\ref{fig:onc_radial_extscl}). When we use the combination of both scaling and artificial extinction (see the radial profiles in the bottom middle panel of Fig.~\ref{fig:onc_radial_extscl}), however, the primordially mass-segregated model can describe the ONC: its $p$-value is above 0.05 between approximately 1.5\,Myr and 3\,Myr (see Fig.~\ref{fig:onc_ks_extscl}). The non-segregated model is still incompatible with the observational data even with all these modifications:  its $p$-value is virtually zero.

Although this last result is not conclusive because the scaling is not dynamically self-consistent, it is consistent with the results of Sect.~\ref{sec:mass_seg}:
the time when the degree of mass segregation in this model is expected to become the same in the primordially mass-segregated and the initially not mass-segregated case is about 8.4\,Myr.
Although our models are also idealistic without a gaseous potential or other realistic features that very young star clusters tend to have \citep{pleiades,subr}, these results are consistent with those reached by \citet{bon_dav98}, who concluded that the ONC most likely formed segregated by mass.

\begin{figure}
        \centering
        \includegraphics[width=\linewidth]{{{onc_ks_new_m5_2.5}}}
        \caption{ $p$-value of the two-sample KS test between the projected cumulative radial distribution of our models (based on ten realisations) and the real ONC.}
        \label{fig:onc_ks}
        
        \vspace*{\floatsep}

        \includegraphics[width=\linewidth]{{{onc_ks_ext_m5_2.5}}}\\
        \vspace*{10pt}
        \includegraphics[width=\linewidth]{{{onc_ks_scl_m5_2.5}}}\\
        \vspace*{10pt}
        \includegraphics[width=\linewidth]{{{onc_ks_extscl_m5_2.5}}}
        \caption{Same as in Fig.~\ref{fig:onc_ks}, but for the scaled model with artificial extinction (top panel), with scaling (middle panel), and with both scaling and extinction. When scaling is involved, time is given in the scaled units ($t_\mathrm{scl} = s_t\, t$), which are valid for all high- and low-mass stars above 0.5\,pc.}
        \label{fig:onc_ks_extscl}
\end{figure}

\begin{figure*}
        \centering
        \includegraphics[width=.82\linewidth]{{{cum_rad_merged_avg_new_mass_2.5}}}
        \caption{Comparison of the average radial profiles (based on ten realisations of our models with 2.4k stars) of the projected models of segregated (blue) and non-segregated clusters (red) with the real ONC (black). In several time frames, we show two mass groups: $m\geq 5\,\msun$ (solid line) and $m<5\,\msun$ (dashed line). The corresponding statistic for each group can be found in Fig.~\ref{fig:onc_ks}, with the comparison even beyond the age of the ONC.}
        \label{fig:onc_radial}
        
        \vspace*{\floatsep}

        \includegraphics[width=.82\linewidth]{{{cum_rad_merged_avg_ext_mass_2.5}}}\\
        \vspace*{10pt}
        \includegraphics[width=.82\linewidth]{{{cum_rad_merged_avg_scl_mass_2.5}}}\\
        \vspace*{10pt}
        \includegraphics[width=.82\linewidth]{{{cum_rad_merged_avg_extscl_mass_2.5}}}
        \caption{Same as in Fig.~\ref{fig:onc_radial}, but for the model with artificial extinction (top row), with scaling of the low-mass stars along the $y$-axis (middle row) and with both scaling and extinction (bottom row).}
        \label{fig:onc_radial_extscl}
\end{figure*}

\section{Conclusions}

Current existing data appear to suggest that embedded clusters may form completely mass segregated. More observational data of newborn and young star clusters are, however, needed to further establish this.

Using idealised numerical $N$-body models of star clusters with 1.2k, 2.4k, 4.7k, and 9.2k stars, we have found that the primordially fully mass-segregated models initially evolve faster (within the first several crossing times). During (or shortly after) the core collapse, the most massive stars (i.e.\ $>5\,\msun$ and even some stars with masses between 2~and $5\,\msun$) loose all signs of primordial mass segregation. For the stars with decreasing mass, more time is needed to balance the degree of mass segregation between the primordially fully mass-segregated and non-segregated models. Using the radial distribution of mean mass in our models, we fitted the time when the primordial mass segregation vanishes, and such a model can no longer be observationally distinguished from an initially not mass-segregated model. The mean time is
\begin{equation*}
        \langle\tauv\rangle \approx 3.3\,\trel \,.
\end{equation*}
In the lower mass models a marginal difference can be still seen after $\tauv$ , but more massive clusters are completely indistinguishable from this point onward.

We cross-matched existing observational data of the ONC from the past 30 years and more from the X-ray, \hbox{(near-)IR,} and optical wavelengths, and we compiled the most complete dataset of this cluster so far. It contains about 2400 sources from brown dwarfs and protostars to high-mass Trapezium stars.
The numerical models with 2.4k stars were then used to study the degree of primordial mass segregation in the ONC. As the high-mass stars ($>5\,\msun$) evolve faster dynamically, their radial profiles quickly reach the observational data in this mass range. This holds for both models. However, neither model can describe the population of low-mass stars that we see in the ONC with a high confidence level (the $p$-value is virtually zero). Two factors were therefore considered in order to approach the observed ONC: (i) extinction and (ii) elongation, as documented in previous works. Although large extinction areas are present in the ONC, including it in the model has only a minor influence on the results and both models are still inconsistent with the data. With a geometrical modification of the modelled clusters according to the observed shape, however, the initially mass-segregated model fits the ONC significantly better at $1.5 \lesssim t \lesssim 3\,\myr$, which is about the current age of the ONC, even in the low-mass range (with $p>0.05$). The initially not mass-segregated cluster is still incompatible with the observations. This result is expected because the estimated time when the 2.4k cluster should forget its primordial mass segregation is about 8.4\,Myr, that is,\ well above the current age of the ONC. Thus, while not conclusive, the data suggest that the ONC may have formed fully segregated by mass. More theoretical and observational study of this intriguing possibility is needed. In the future, more detailed models that consider the evolution of non-spherical star clusters should be analysed. Other models of the ONC that take the three putative populations \citep{beccari_pop} as well as a high initial binary population \citep{kroupa95,pleiades,belloni17} into account should also be constructed.

\begin{acknowledgements}
This study was supported by Charles University through grants GAUK-186216 and SVV-260441.
We sincerely thank the participants and organisers of the ``M+3rd Aarseth N-Body Meeting'' for a fruitful discussion about this topic.
We also greatly appreciate access to the computing and storage facilities owned by parties and projects contributing to the National Grid Infrastructure MetaCentrum, provided under the programme ``Projects of Large Research, Development, and Innovations Infrastructures'' (CESNET LM2015042).
\end{acknowledgements}

\bibliographystyle{aa}
\bibliography{ONC_mass_segr}

\clearpage
\appendix
\section{Information on the initial state of the clusters}
\label{ap:models}

All the models (1.2k, 2.4k, 4.7k, and 9.2k) were generated from the same distribution function of positions and velocities, IMF, and with the same two degrees of mass segregation according to \citet{baumgardt_segr}. Therefore, only the most populous model with 9.2k stars is used for a further discussion of the initial conditions. As our models do not contain any primordial binary stars, the evolution towards thermal equilibrium due to the energy equipartition process is dominated by random two-body encounters.

\begin{figure*}
        \centering
        \includegraphics[width=\linewidth]{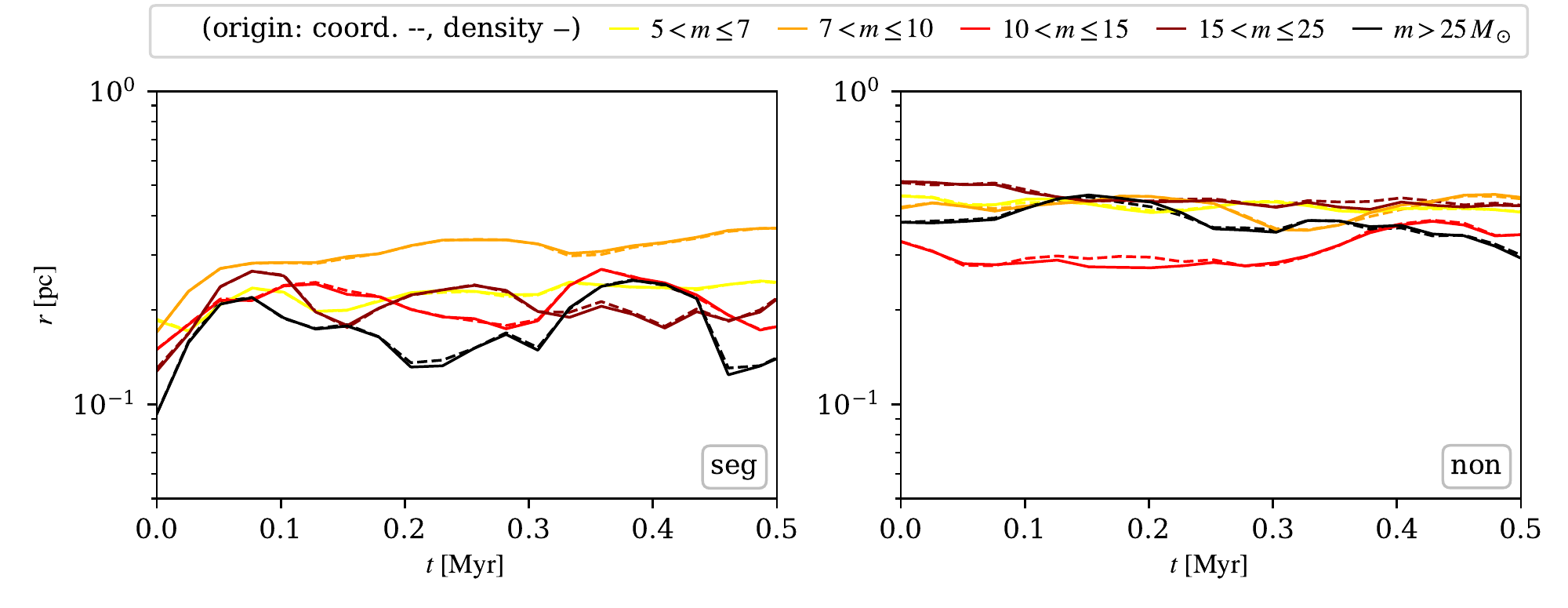}
        \caption{Evolution of the radii of high-mass stars in time (up to 0.5\,Myr, i.e.\ approximately $3.5\,\tcr$) of the 9.2k model (one realisation each of \mseg\ and \mnon\ is plotted). Colours represent stars in different mass bins (from $5\,\msun < m \leq 7\,\msun$ up to $m>25\,\msun$). The bins contain 39, 26, 18, 13, and 11 stars, respectively. The mean position within each bin is given. Radii with respect to the coordinate origin (dashed lines) and the density centre (solid lines) are plotted.}
        \label{fig:radius}
\end{figure*}

A higher concentration of high-mass stars in the core of the initially mass-segregated cluster allows for sooner and more energetic interactions between them and consequently for a sooner expansion of the central region than in the initially not mass-segregated cluster. This property is discussed in Sect.~\ref{sec:mass_seg} to explain the greater expansion of the 1.2k \mseg\ model. The early core expansion of the \mseg\ model is also visible in Fig.~\ref{fig:radius}, where we plot the mean positions of stars in five mass bins above $5\,\msun$. In comparison to the \mnon\ cluster, where the radii are almost stationary, the \mseg\ cluster stars move more abruptly. In this figure, we plot not only the radii with respect to the density centre (see Sect.~\ref{sec:models}), but also with respect to the rigid Cartesian coordinate system predefined at the beginning of each integration. For each bin, the two curves coincide. Thus, we conclude that this effect is dynamical and not produced by our choice of the coordinate system.

\begin{figure*}
        \centering
        \includegraphics[width=\linewidth]{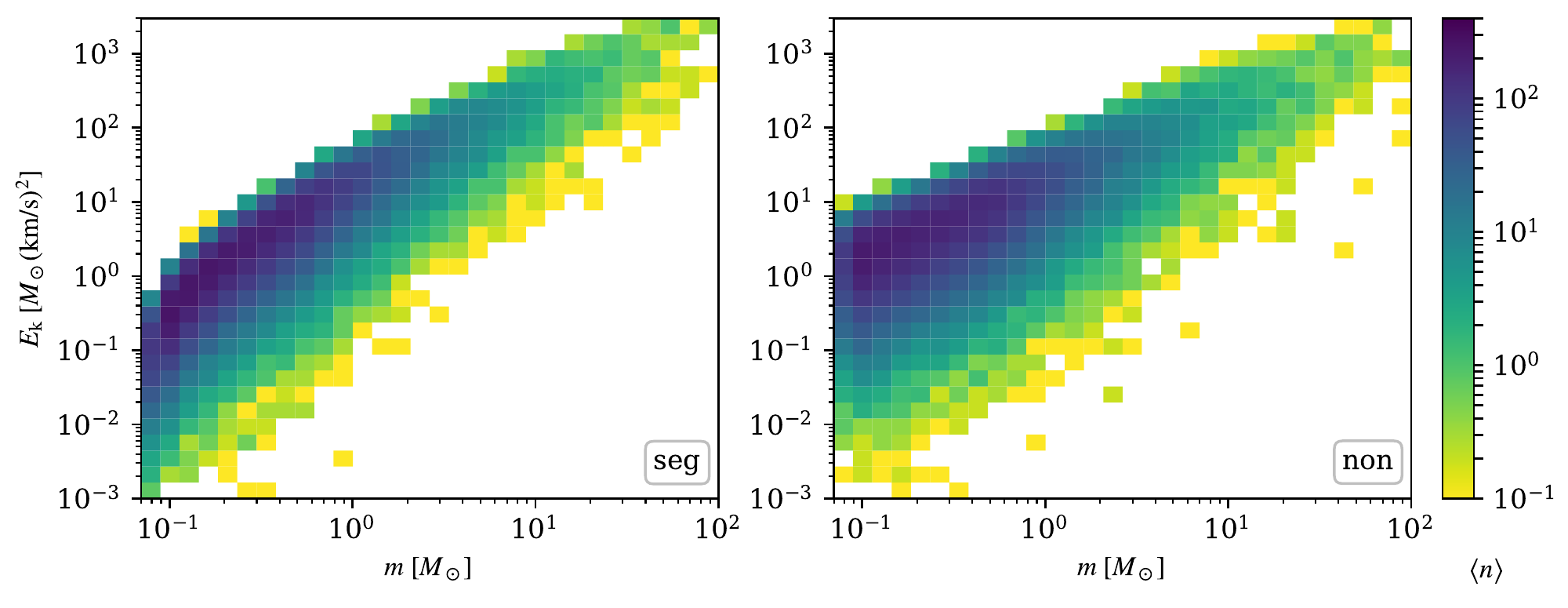}\\
        \includegraphics[width=\linewidth]{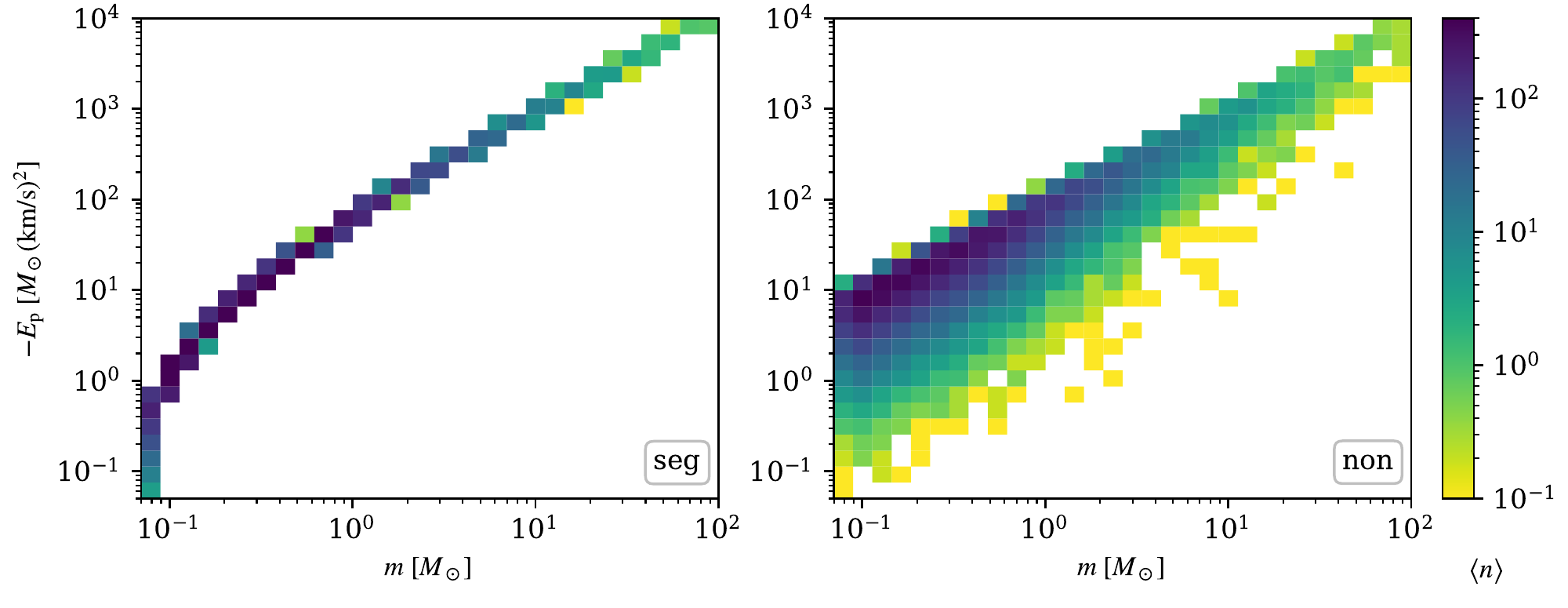}
        \caption{Initial distribution of stellar kinetic energy (top panels) and potential energy (bottom panels) with respect to mass in the 9.2k model for both primordially mass-segregated and non-segregated clusters. The colour scale represents the mean number of stars based on ten realisations of this model.}
        \label{fig:kin_pot}
\end{figure*}

The initial conditions of both mass-segregated and non-segregated models are also plotted in Fig.~\ref{fig:kin_pot}, which shows the dependence of kinetic and potential energy on the masses of stars. By definition, high-mass stars of the \mseg\ model are concentrated more strongly than in the \mnon\ model; as is visible in the lower panels of this figure, the stars in the \mseg\ model are arranged perfectly by mass and potential energy (the highest masses have the lowest potential energy), while in the \mnon\ model we see a scatter. The distribution of kinetic energy (in the top panels) in both \mseg\ and \mnon\ models is similar, except for the very low-mass range. This shows that the high-mass stars that are localised in the core in the initially mass-segregated model are not overcolled, even though there are no primordial binaries. This in turn means that the most massive component may expand immediately.

\clearpage
\section{Additional figures: Mass segregation}
\label{ap:segr}

\begin{figure}
        \centering
        \includegraphics[width=\linewidth]{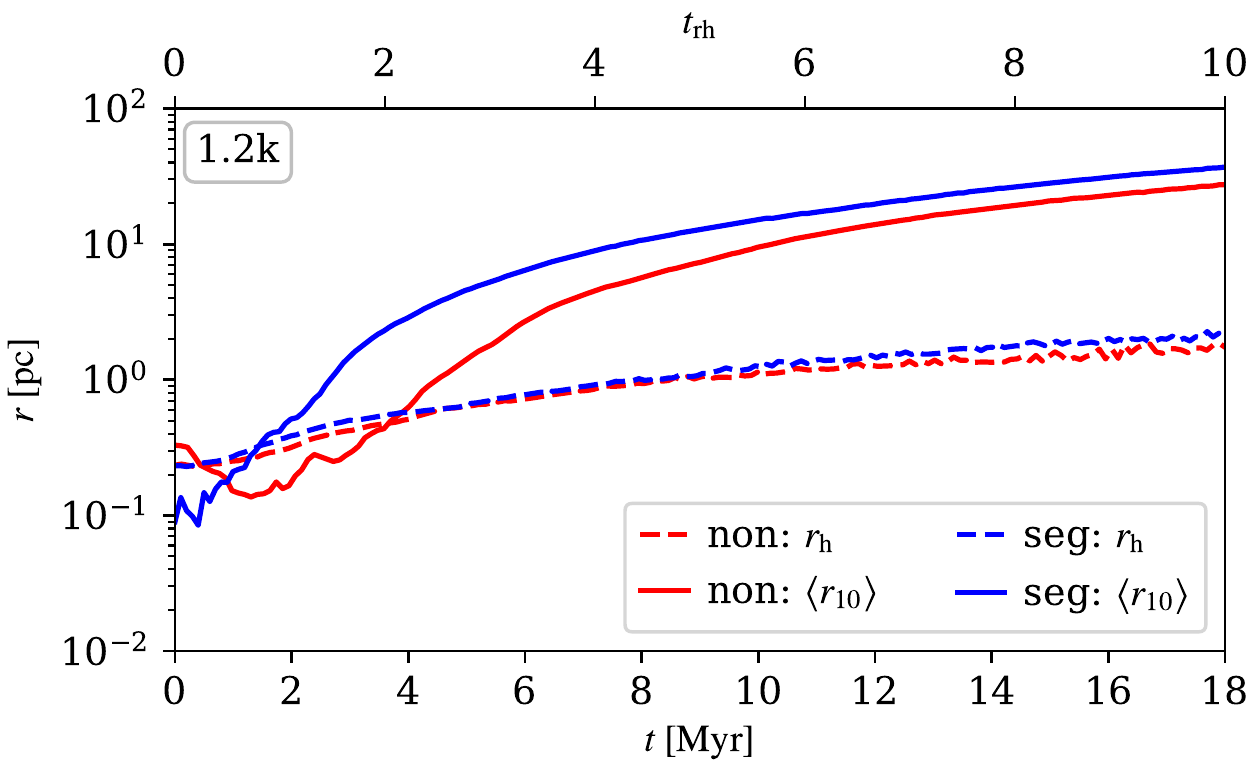}\\
        \includegraphics[width=\linewidth]{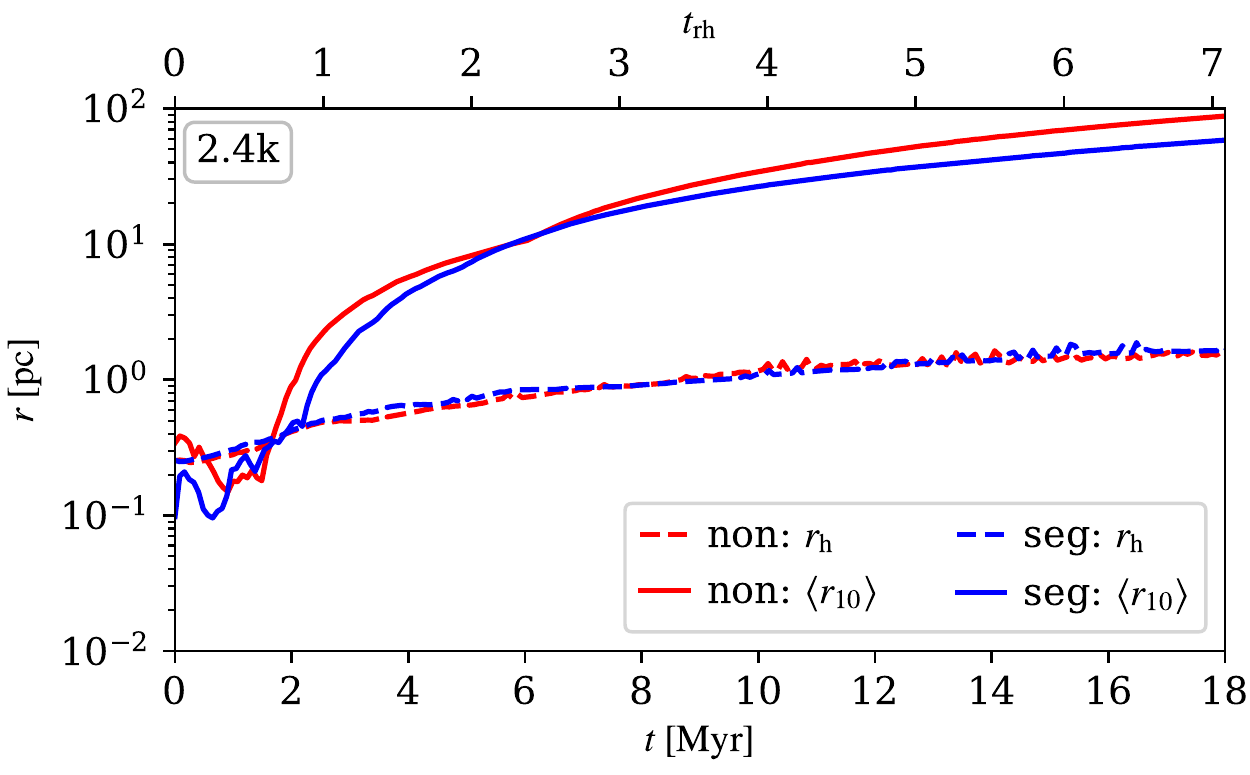}\\
        \includegraphics[width=\linewidth]{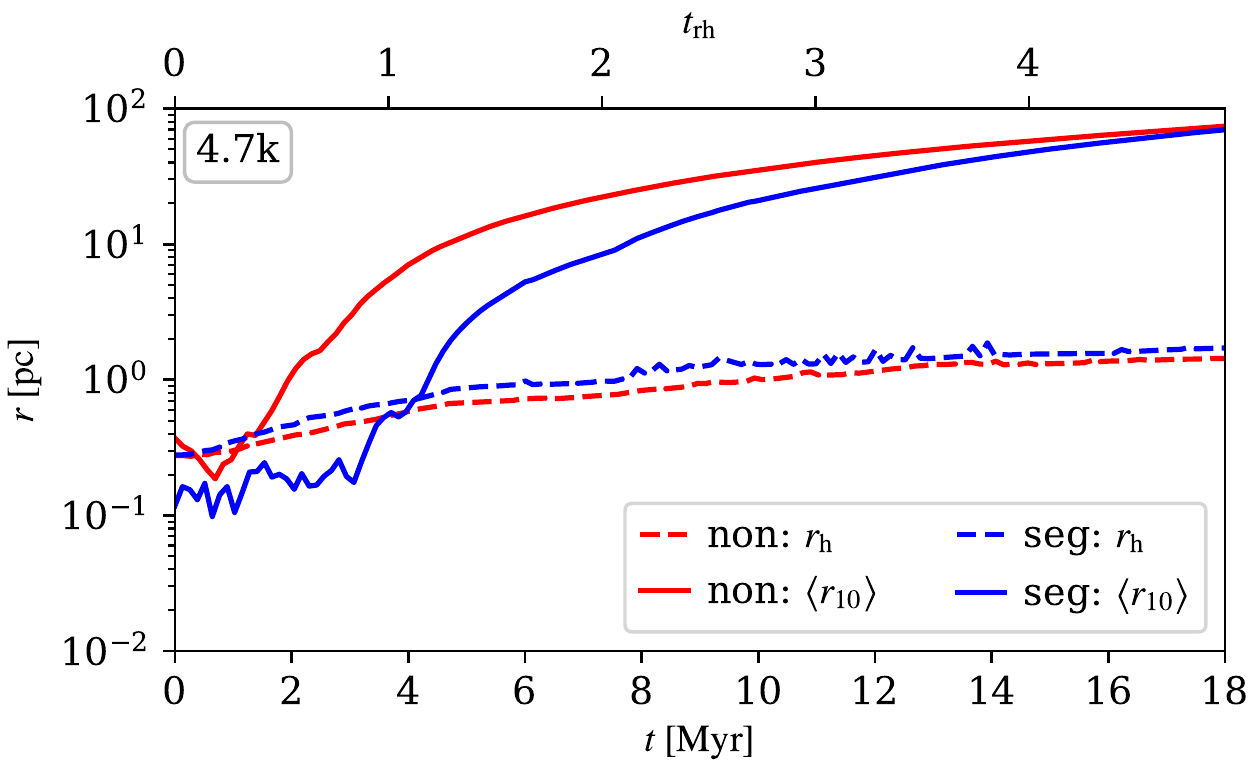}\\
        \includegraphics[width=\linewidth]{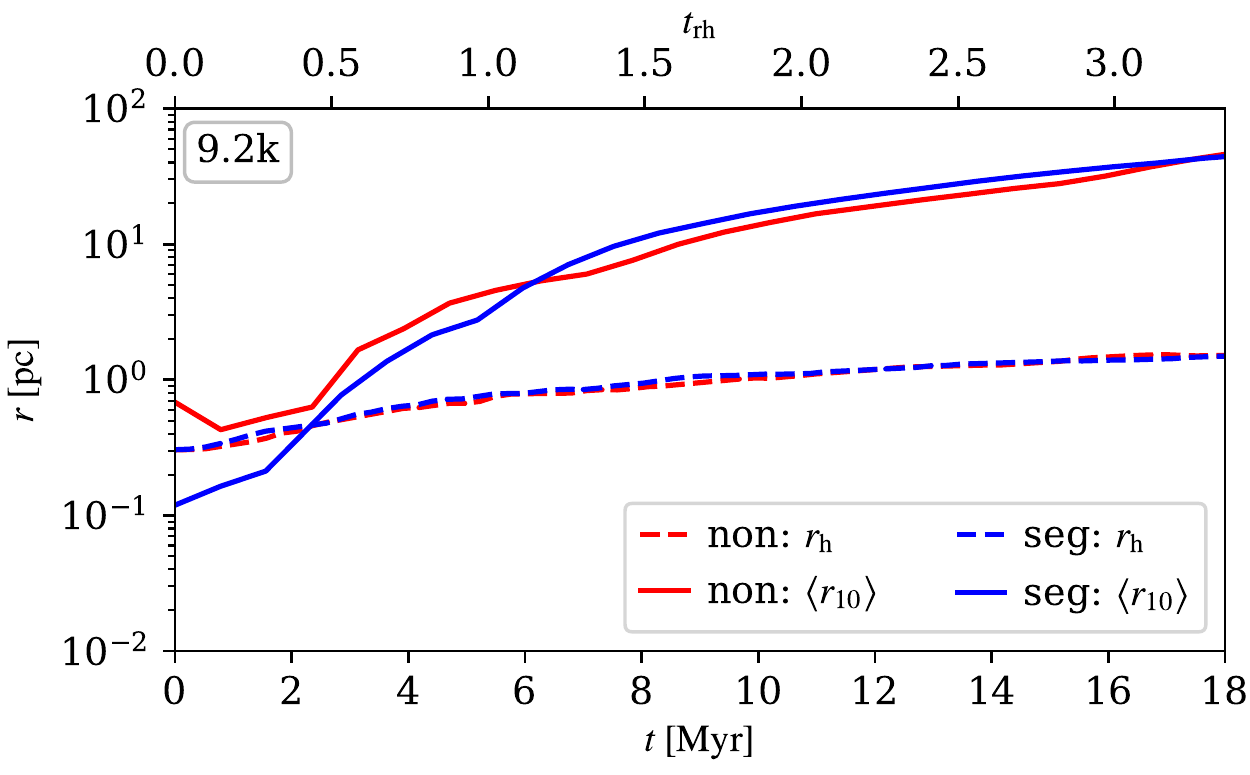}
        \caption{Average positions of the most massive stars that are in the top 10\,\% of the total mass of the cluster (solid lines). Models with 1.2k, 2.4k, 4.7k, and 9.2k stars are plotted from top to bottom. The data are averaged over all realisations of a given model. For reference, we also plot the half-mass radius with a dashed line. Time extends to ${\approx}18\,\myr$.}
        \label{fig:top_mass}
\end{figure}

We further discuss mass segregation through the distribution of stars of different masses within each model. In Fig.~\ref{fig:top_mass} we include the mean position of the most massive stars in each model whose masses add up to 10\,\% of the total mass. These curves are also averaged over all realisations of the model (the stellar masses were optimally sampled from the IMF in each realisation, so that we may average the positions of individual stars from different realisations). We note that after the core collapse, the most massive stars from both \mseg\ and \mnon\ clusters expand at a very similar rate. In the smallest clusters, we can still see some difference in the rate of expansion of these stars, but with an increasing total mass of the model, any visual difference linked to the degree of primordial mass segregation disappears.

\begin{figure*}[h]
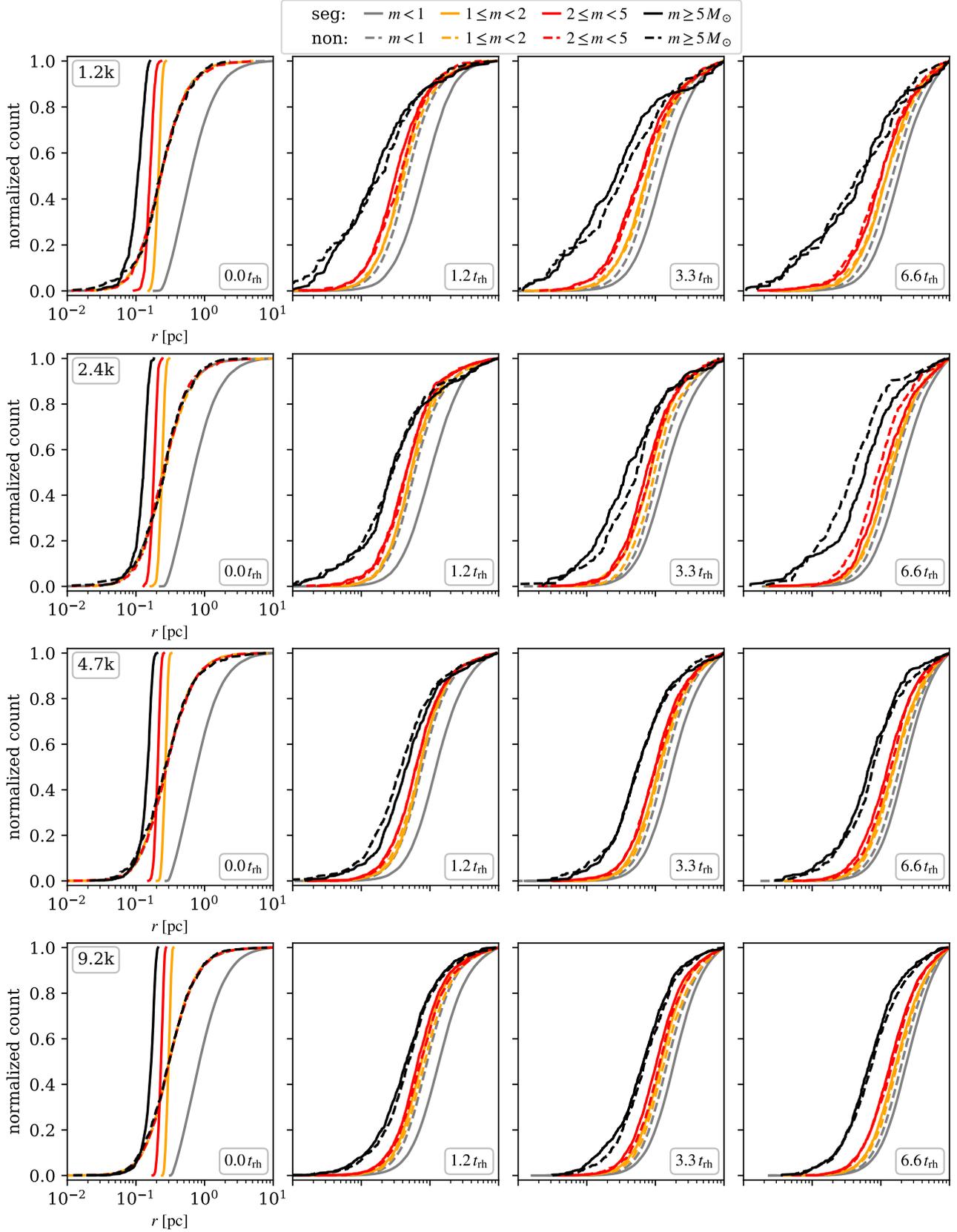

        \centering
        \includegraphics[width=.925\linewidth]{{{m659sun_onc_R_mall_10.0}}}\\
        \includegraphics[width=.925\linewidth]{{{m1318sun_onc_R_mall_10.0}}}\\
        \includegraphics[width=.925\linewidth]{{{m2636sun_onc_R_mall_10.0}}}\\
        \includegraphics[width=.925\linewidth]{{{m5272sun_onc_R_mall_10.0}}}
        \caption{Cumulative distribution functions of stars in four different mass ranges in several key time frames (the initial conditions, $t\approx 0.5\,\tauv$, $t\approx \tauv$ and $t\approx 2\,\tauv$, given in the units of the median relaxation time, see also Fig.~\ref{fig:tau_v}). Models with 1.2k, 2.4k, 4.7k, and 9.2k stars are plotted (from top to bottom).}
        \label{fig:cumulative}
\end{figure*}

The cumulative radial profiles that are plotted in Fig.~\ref{fig:cumulative} provide other evidence that the initial mass segregation is gradually lost. Initially very distinct curves start to look similar; this occurs first for the most massive stars. At half a median relaxation time, that is,\ after the core collapse, there is no difference between the distribution of high-mass stars ($> 5\,\msun$), and this seems to also be the case even in the mass range between 2~to $5\,\msun$. The lower the masses of the stars, the longer before their radial distributions of the initially mass-segregated and not mass-segregated clusters coincide. This process becomes faster with higher total mass of the model.

Although the curves plotted in Fig.~\ref{fig:cumulative} show a slight difference, the $p$-value calculated from a two-sample KS test between the same mass groups from the \mseg\ and \mnon\ models is higher than 0.4 (sometimes exceeding even 0.9) in the mass groups with $m>1\,\msun$ by the time we reach $\tauv$; the mass groups with $m>5\,\msun$ of the two models start to coincide even sooner and are completely indistinguishable at approximately $0.3\,\tauv$. We note that for $p>0.05,$ the hypothesis that two samples come from the same distribution cannot be excluded with more than 5\,\% confidence. The only problem arises in the low-mass range with $m<1\,\msun$ , where the $p$-value is much more dependent on the range of radii used for the KS test than in the other mass groups, especially the subgroup of masses $m<0.5\,\msun$. When we include the radii only up to 5\,pc, the \textit{p}-value exceeds the critical value at about the time of the third panel (i.e.\ $\tauv$). When we include stars up to 10\,pc, it becomes lower and oscillates around the critical value of 0.05. Nevertheless, given the spacial distributions of high-mass stars that dominate mass segregation, we conclude that the primordial mass segregation is lost by the time $\tauv$.

\clearpage

\clearpage
\section{Bin-weighting of parameter $A$}
\label{ap:bins}

\begin{figure}
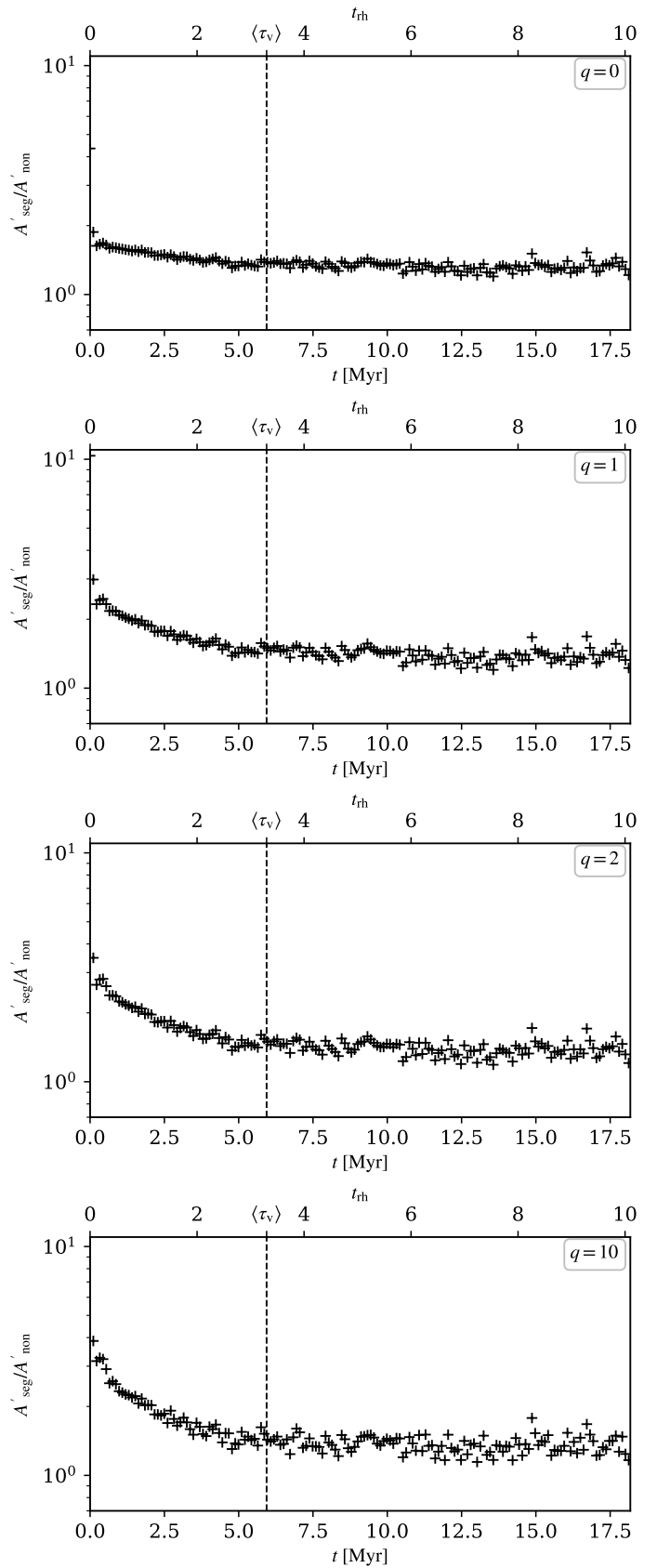

        \centering
        \includegraphics[width=\linewidth]{{{m659sun_seg_par_bin_p0_0.1-10.0}}}\\
        \includegraphics[width=\linewidth]{{{m659sun_seg_par_bin_p1_0.1-10.0}}}\\
        \includegraphics[width=\linewidth]{{{m659sun_seg_par_bin_p2_0.1-10.0}}}\\
        \includegraphics[width=\linewidth]{{{m659sun_seg_par_bin_p10_0.1-10.0}}}
        \caption{Comparison of different values of the weight power index $q$ from Eq.~\eqref{eq:area_general}, which is set to 0, 1, 2, and 10 (from top to bottom). We compare only the data from the 1.2k model as an example. The mean time $\langle\tauv\rangle$ (see Eq.~\eqref{eq:tauv}) is also plotted for reference. The limits of the vertical axes are fixed.}
        \label{fig:weights}
\end{figure}

In Sect.~\ref{sec:mass_seg} we have defined a way of measuring the difference of mass segregation between models of the same initial mass and initial radius by the means of the bin-weighted sum of histogram heights, see Eq.~\eqref{eq:area}. We chose to weight each bin by its inverse width $\Delta r_k$ (equidistant in the logarithmic scale) to emphasise the importance of mass segregation, which is visible especially at small radii. Generally, the weight $\Delta r_k$ may be raised to a non-negative power index $q$,
\begin{equation}
        \label{eq:area_general}
        A' = \sum_{k = 1}^{\nbin}{\frac{\langle m(r_k) \rangle}{(\Delta r_k)^q}} \,,
\end{equation}
compare with Eq.~\eqref{eq:area}.
In Fig.~\ref{fig:weights} we plot the ratio $A'_\mseg / A'_\mnon$ for the 1.2k model as an example to show how the choice of $q$ influences the results (we set $q=0$, 1, 2, and 10).

With an increasing index $q$, the initially large difference is increased and also renders the mass segregation decrease more visible. It is also evident that with a larger $q$ the fluctuations in increase, $A'_\mseg / A'_\mnon$, especially after $\langle\tauv\rangle$. Nevertheless, the time $\langle\tauv\rangle$, when the information of the primordial mass segregation is forgotten, is invariant on the value of $q$.
The choice of $q=1$, as we presented in Sect~\ref{sec:mass_seg}, represents the first order. It was chosen arbitrarily, but the results do not loose generality.

We note that in order to visualise differences between the values of $q$ and the shape of $A'_\mseg / A'_\mnon$ in Fig.~\ref{fig:weights}, the limits of the vertical axes were set to the same values. Therefore, the data point at $t=0$, which symbolises the large initial difference between the mass-segregated and non-segregated model (e.g.\ see the panels for $t=0$ in Fig.~\ref{fig:mean_mass}), is visible only in the plots of $q=0$ and $q=1$ but is outside the range of the plots of $q=2$ and $q=10$.

In the discussion of the lower mass models, we stated that the initially mass-segregated cluster inflates more than the non-segregated one. Therefore the ratio of $A'_\mseg / A'_\mnon$ stays at a slightly higher value than 1. It might seem reasonable to try to lower this effect, for example, by normalising the radial distribution of the mean mass (in Fig.~\ref{fig:mean_mass}) by the current half-mass radius. For this method of comparing the mean masses to work, however, the bins must remain equally spaced throughout the whole cluster evolution. Because the half-mass radius evolves, we may not use it for normalisation. In principle, we could scale both models to their initial half-mass radius, but as its value for the primordially fully mass-segregated and non-segregated cluster is similar, we may also chose any other fixed units, for instance,\ parsecs. However, if we were about to compare clusters of different initial sizes, we would have to scale both to the same size for the method to work (e.g.\ again the initial half-mass radius).

\end{document}